\documentclass[pdflatex,sn-mathphys-num]{sn-jnl}

\usepackage{graphicx}%
\usepackage{multirow}%
\usepackage{amsmath,amssymb,amsfonts}%
\usepackage{amsthm}%
\usepackage{mathrsfs}%
\usepackage[title]{appendix}%
\usepackage{xcolor}%
\usepackage{textcomp}%
\usepackage{manyfoot}%
\usepackage{booktabs}%
\usepackage{algorithm}%
\usepackage{algorithmicx}%
\usepackage{algpseudocode}%
\usepackage{listings}%
\usepackage{subcaption} 
\usepackage{xfrac} 

\theoremstyle{thmstyleone}%
\theoremstyle{thmstyletwo}%

\theoremstyle{thmstylethree}%

\begin{document}

\title[Article Title]{The Effect of Flow Parameters and Wall Models on Gas-Surface Interactions: A Numerical Investigation of dsmcFoam} 

\author*[1]{\fnm{M. Burak} \sur{Agir}}\email{burak.agir@manchester.ac.uk}\equalcont{ORCID: 0000-0001-7500-6376} 

\author[1]{\fnm{Nicholas H.} \sur{Crisp}}

\author[1]{\fnm{Katherine L.} \sur{Smith}}

\author[1]{\fnm{Peter C. E.} \sur{Roberts}}

\author[2]{\fnm{Mike} \sur{Newsam}}

\author[2]{\fnm{Matthew} \sur{Griffiths}}

\author[3]{\fnm{Shreepali} \sur{Vaidya}}

\affil[1]{\orgdiv{Space Systems Engineering Research Group}, \orgname{The University of Manchester}, \orgaddress{\street{Oxford Road}, \city{Manchester}, \postcode{M13 9PL}, \country{United Kingdom}}}

\affil[2]{\orgname{Stellar Advanced Concepts Ltd}, \orgaddress{\street{Translation and Innovation Hub, Imperial College White City Campus}, \city{London}, \postcode{W12 0BZ}, \country{United Kingdom}}}

\affil[3]{\orgname{Stellar Space Industries B.V.}, \orgaddress{\street{`s-Gravendijckseweg 31}, \city{Noordwijk}, \postcode{2201 CZ}, \country{The Netherlands}}}

\abstract{Atmosphere-breathing electric propulsion systems harness atmospheric particles as propellant, enabling efficient operation across diverse environmental conditions. To accurately simulate the captured gas flow through the modules, particle-surface interactions must be carefully modelled. To initiate this research, a parametric study is conducted using an extensive simulation matrix to investigate the effects of flow parameters, such as velocity, temperature, species, and angle of attack, and wall model parameters (diffuse fraction/accommodation coefficient) on gas-surface interactions. A simplified test geometry was created to run 2D simulations, where the flow interacts with an adjacent wall positioned perpendicular to one of the inlet patches. In this study, changes in reflection patterns, force density on the surface, and flow properties in the vicinity of the wall are investigated under varying flow and wall conditions using the current boundary conditions of the dsmcFoam solver. Furthermore, the capabilities of dsmcFoam's default boundary conditions in predicting gas-surface interaction physics are evaluated using the results of the simulation matrix. The findings highlight the need for new boundary conditions to accurately replicate interaction physics across various aspects. } 

\keywords{Gas-surface Interactions, direct simulation Monte Carlo, dsmcFoam, Atmosphere Breathing Electric Propulsion}

\maketitle

\section{Introduction}\label{sec:introduction}
	The power budget of Atmosphere-Breathing Electric Propulsion (ABEP) systems is closely tied to their ability to compensate for drag. Intake performance is crucial for improving propellant capture and contribute to drag compensation. As the flow travels through the modules of an ABEP system, the flow regime shifts from free molecular to slip by compression, causing gas-surface interactions to vary throughout the particle collection and compression process. Although particle collisions are rare in the highly rarefied free stream, wall interactions strongly influence the parameters of the highly collimated flow between the intake and compressor modules. The need for an advanced wall model has emerged to ensure accurate end-to-end simulations and achieve convergence that closely aligns with realistic physics.
	
	The interaction between gas particles and the surface plays a pivotal role in determining surface forces and flow patterns around the interaction surface. Numerically, these interactions are modeled by several factors, including the choice of wall model, which governs the particle reflection kernel and may consider the energy and momentum exchange at the boundary, i.e., interaction wall, and the flow parameters, such as flow distortion in the vicinity of the wall, velocity and temperature of the flow, which is a mixture of reflected and incoming particles. 
    	
	Several studies in the literature have used dsmcFoam+, a previous version of dsmcFoam, for the simulation and optimisation of the passive intake module of ABEP systems. Rapisarda \textit{et al.}~\cite{rapisarda2023design} discussed various designs for the intake module and their optimisation using the Maxwellian wall model in dsmcFoam+, which is a mixture of specular and diffuse wall models. The authors emphasise that this wall model presents a simplified approach to the complex phenomenon of gas-surface interactions, as experiments show that gas-surface interactions do not exhibit a completely specular or diffuse scattering kernel. Rapisarda~\cite{rapisarda2023modelling} also conducted a study simulating the intake module using dsmcFoam+, where the intake walls were considered fully diffuse. A recent study~\cite{andrews2024cathode} also modelled the intake walls as fully diffuse. Romano \textit{et al.}~\cite{romano2021intake} employed the PICLas computing code for intake design, and two wall models were selected: fully specular and fully diffuse walls. As stated in Ref.~\cite{romano2021intake}, several studies have developed intake modules with diffuse scattering materials and small ducts to tailor the flow and trap molecular flows to increase efficiency. For instance, Refs.\cite{romano2015air,romano2016intake} conducted DSMC simulations using diffuse walls for intake design investigations; however, fully specular wall models were also employed in some studies~\cite{jackson2018conceptual}. However, the realistic scattering kernel should lie between the extrema of diffuse, specular, and isotropic~\cite{struchtrup2013maxwell}.
	
	While existing models in the dsmcFoam computing code provide some insights, a converged effect of material properties on gas-surface interactions for diverse species and varying environmental factors is needed. This limitation hinders accurate simulation of flow physics, especially when the flow transitions through different regimes within ABEP modules.
	
	To address these challenges, this study aims to assess the effectiveness of existing wall models in dsmcFoam under a range of flow conditions relevant to ABEP applications. The findings will help identify key requirements for developing a more robust and realistic wall model to enhance the predictive capabilities of numerical simulations for these systems.
	
	In this study, the numerical wall models currently available in dsmcFoam are tested with varying flow parameters, such as angle of attack (AoA), free-stream velocity, and free-stream temperature, which affect Knudsen number (Kn) while maintaining constant pressure, and species of the working gas. Thus, the demands for a more comprehensive wall model for ABEP applications will be investigated. 

\section{Methodology}\label{sec:methodology} 
\subsection{Numerical Technique}\label{subsec:numericalTechnique} 
In gas flows, as the mean free path —the average distance a molecule travels before colliding with another gas molecule~\cite{white_thesis}— increases, the flow becomes more rarefied, leading to a breakdown in continuum flow conditions. The transport terms in the Navier-Stokes-Fourier (NSF) equations rely on macroscopic flow variations, therefore, the reduction in density disrupts continuum assumptions and the linear constitutive relations for heat flux and shear stress, causing these equations to break down. As a result, the NSF equations are no longer applicable for accurately modelling rarefied gas flows~\cite{bird,Gad-el-Hak}. 
	
	Direct simulation Monte Carlo (DSMC) is a method developed by Bird~\cite{bird} to simulate rarefied gas flow at the kinetic scale. This stochastic method provides accurate results even for highly rarefied flows~\cite{garcia}. With this applicability over a wide range of rarefaction levels, the DSMC molecular scheme is now widely used by many communities to seek solutions for rarefied gas dynamics problems. Over the years, different DSMC solvers have been developed to address different challenges. 
	
	The degree of rarefaction and flow regime of a gas are determined by Kn, defined as $Kn=\lambda/L$, where $\lambda$ is the mean free path and \textit{L} is a characteristic length. The characteristic flow dimension represents the scale length of the macroscopic flow quantity gradients~\cite{bird_annu.rev.} when calculating the local Kn; therefore $Q$ can be the temperature, density, velocity or pressure~\cite{oran}; e.g. $L= Q/|\nabla Q|$. 
		
	The flow regimes can be classified according to the value of Kn. As Kn approaches zero, the interactions between the molecules (intermolecular collisions) increase, and the gas exhibits excellent thermal equilibrium.  As the value of Kn approaches infinity, the degree of rarefaction increases, and intermolecular interactions become fewer. 

\subsection{Computing Code}\label{subsec:computingCode} 
dsmcFoam (previously dsmcFoam+) is an open-source solver for rarefied gas problems, which is implemented within OpenFOAM (or \textbf{Open}-source \textbf{F}ield \textbf{O}peration \textbf{A}nd \textbf{M}anipulation). The OpenFOAM suite~\footnote{OpenFOAM Website. \url{https://www.openfoam.com}. Accessed on 07/11/2024.} offers extensive pre-processing capabilities and widespread adoption, with its GPL3 licensing ensuring accessibility of ongoing developments for a broad user base. Consequently, the availability of a DSMC solver within the robust OpenFOAM framework grants users and developers access to a variety of well-established features. The software is capable of modelling steady-state and transient simulations, different molecular models such as variable hard sphere (VHS) and variable soft sphere (VSS) collisions, no-time-counter (NTC) collision scheme, etc.~\cite{scanlon_dsmcFOAM}. 
	
	The first version of dsmcFoam was introduced in 2010~\cite{scanlon_dsmcFOAM}, followed by the enhanced version dsmcFoam+ in 2018~\cite{white}, which has since been augmented with numerous additional features. Since then, the solver has been widely used for various rarefied gas applications, including ABEP intake design and simulation. A cross-model validation was performed in Ref.~\cite{rapisarda2023modelling} using dsmcFoam+ to assess the performance of the Atmosphere-Breathing Ion Engine (ABIE) intake, and a study on the design and optimisation of the intake module of an ABEP system was conducted in Ref.~\cite{rapisarda2023design}, also using dsmcFoam+. 
	
	The design and simulation of ABEP intake modules require modeling across flow regimes that transition from free-molecular to slip during the pre-air-breathing propulsion phase (throughout the intake and compressor modules). This necessitates adjustments in time-step and mesh size to accurately capture intermolecular collisions for a comprehensive end-to-end simulation. In response to this need, recent updates in OpenFOAM-v2406 have incorporated dynamic adaptation capabilities into dsmcFoam. These advancements include the automated adjustment of time-step and sub-cell. This dynamic feature enables dsmcFoam to handle transitions in flow regimes autonomously. 
    
	\section{Boundary Conditions in dsmcFoam} 
	Once the geometry of a control volume is created, the boundaries of the defined control volume should be correctly linked to the appropriate \textit{boundaryModels}, which specify the selected boundary functions. The boundary model is linked to the \textit{patch} to represent the properties of the wall. In this study, simulations are performed using four boundary conditions:  
	
	\subsection{Specular Wall Model: \textit{dsmcSpecularWallPatch}}
	The gas particles collide with a perfectly smooth and rigid wall at a velocity and reflect back by inverting the normal component of the velocity while the tangential velocity remains unchanged. The reflection kernel is a delta function, 
	
	\begin{equation}
	R_{spec} (c^{\prime}_{k} \rightarrow c_{k}) = \delta(c^{\prime}_{k} - c_{k} + 2n_{j}c_{j}n_{k}), 
	\end{equation}
	
	\noindent 
	where $c^{\prime}_{k}$ and $c_{k}$ represent the reflected velocity component in the \textit{k}-th direction after interaction with the wall and the incoming velocity component in the \textit{k}-th direction before interaction with the wall, respectively. The $\delta$ function enforces exact conservation of velocity, ensuring that the reflected velocity satisfies the specular condition. Here, $n_{j}$ is the \textit{j}-th component of the unit normal vector (\textbf{n}) at the wall surface, $n_{k}$ is the \textit{k}-th component of the unit normal vector on the wall surface, and $c_{j}$ is the incoming velocity component in the \textit{j}-th direction~\cite{struchtrup2013maxwell}. 
	
	\subsection{Diffuse Wall Model: \textit{dsmcDiffuseWallPatch}}
	On average, collisions between the gas and the wall tend to drive the gas toward equilibrium with the wall. For a fully diffuse wall, where the diffuse fraction is unity, the gas particles are reflected in a Maxwellian distribution characterised by the wall temperature, $T_{w}$, and the wall velocity, $\vartheta^{W}_{k}$. The diffuse reflection kernel is
	
	\begin{equation}
	R_{diff} (c^{\prime}_{k} \rightarrow c_{k}) = |c_{n}| f_{0}(T_{W}, c_{k}), 
	\end{equation}
	
	\noindent where $f_{0}$ is the Maxwellian distribution of diffusively reflected particles~\cite{struchtrup2013maxwell}. It is significant to highlight that the post-collision terms of the particles are independent of the pre-collision terms in the diffuse reflection kernel in dsmcFoam~\cite{bird}.
	
	In dsmcFoam, the diffusive reflection is completed at the surface temperature. The boundary updates the velocity of the particle and adds the velocity of the boundary if it has any. Finally, the energies of the reflected particle are also updated based on the surface temperature.
	
	In cases where a gas has a velocity component parallel or closely parallel to the surface, the stagnation temperature in the gas differs from the static temperature. Thus, the distribution function for the incoming particles will not be the same as that for the reflected particles or the particles near the surface, which will not be a Maxwellian distribution~\cite{bird}. In the default diffuse boundary condition of dsmcFoam, if the tangential velocity is sufficiently small, a perturbation is added to the incoming velocity, and the velocity is recalculated for reflection. 
	
	\subsection{Maxwellian Wall Model: \textit{dsmcMixedDiffuseSpecularWallPatch}}
	For a realistic reflection kernel, Maxwell suggested a linear combination of fundamental kernels~\cite{maxwell1879vii} with constant factors of $\Gamma_{0}$ for specular, $\Lambda_{0}$ for isotropic, and $\Theta_{0}$ for diffuse reflections as, 
	
	\begin{multline}
	R_{M} (c^{\prime}_{k} \rightarrow c_{k}) = \Theta_{0} R_{diff} (c^{\prime}_{k} \rightarrow c_{k}) + \Gamma_{0} R_{spec} (c^{\prime}_{k} \rightarrow c_{k}) + \Lambda_{0} R_{scat} (c^{\prime}_{k} \rightarrow c_{k}). 
	\end{multline} 
	
	\noindent 
	As individual kernels satisfy the principle of reciprocity, the weighted sum of these kernels also adheres to this property. Normalisation is ensured by requiring that the weighting factors sum to one~\cite{struchtrup2013maxwell}:  
	
	\begin{equation}
	\Gamma_{0} = 1 - \Theta_{0} - \Lambda_{0}. 
	\end{equation}
	
	\noindent 
	In Maxwell's classical model~\cite{maxwell1879vii}, the isotropic reflection is neglected, therefore, $\Lambda_{0} = 0$ and $\Gamma_{0} = 1 - \Theta_{0}$. 
	
	In dsmcFoam, the Maxwell gas-surface interaction model is generated as a mixture of specular and diffuse wall models, where specular reflection represents perfectly smooth surfaces and diffuse reflection is for microscopically rough surfaces. In both reflection kernels, the pre-collision parameters are not considered; however, post-collision parameters are updated depending on the selected reflection kernel for each particle. In this boundary condition, a user-defined diffuse fraction is introduced to simulate the surface more realistically, allowing to define the fraction of particles that interact with the wall in a diffuse manner~\cite{white_thesis}. 
	
	\subsection{Cercignani-Lampis-Lord Wall Model: \textit{dsmcCLLWallPatch}}
	The CLL model was developed based on the scattering kernel proposed by Cercignani and Lampis~\cite{cercignani1971kinetic}, using consistent experimental scattering distribution data. This kernel considers that the tangential and normal components of the reflection velocity have two separate accommodation coefficients: normal, $\alpha_{n}$, and tangential, $\alpha_{t}$. The scattering physics is always a function of the incoming parameters of the particles, and the kernel is 
	
	\begin{multline}
    R_{CL} (c^{\prime}_{k} \rightarrow c_{k}) = \frac{1}{2\pi} \frac{1}{\alpha_{n} \alpha_{t} (2 - \alpha_{t})} \frac{c_{n}}{\big(\frac{k}{m}T_{W}\big)^{2}} \frac{1}{2\pi} \int^{2\pi}_{0} exp \bigg[\frac{\sqrt{1 - \alpha_{n}} c_{n} c^{\prime}_{n} cos\phi_{n}}{\alpha_{n} \frac{k}{m} T_{W}} \bigg] d\phi_{n} \\
	\times exp \bigg[ -\frac{c^{2}_{n} + (1 - \alpha_{n}) {{c^{\prime}}^{2}}_{n}}{2 \frac{k}{m} T_{W} \alpha_{n}} - \frac{c^{2}_{t} - 2(1 - \alpha_{t})\textbf{c}_{t} \cdot \textbf{c}^{\prime}_{t} + (1 - \alpha_{t})^{2} {{c^{\prime}}^{2}}_{t}}{2 \frac{k}{m} T{W} \alpha_{t} (2 - \alpha_{t})}\bigg], 
	\end{multline}
	
	\noindent 
	where $\textbf{c}_{t}$ is the 2D tangential velocity vector~\cite{struchtrup2013maxwell}. 
	
	In dsmcFoam, the scattering kernel proposed in Refs.~\cite{lord1989application,lord1991some} has been implemented, and the model has been extended to include rotational energy exchange at the surface. This boundary condition requires three user-defined accommodation coefficients, which are normal, tangential, and rotational. 

\section{Problem Description}
\begin{figure} [ht]
		\centering
		\includegraphics[width=0.9\columnwidth]{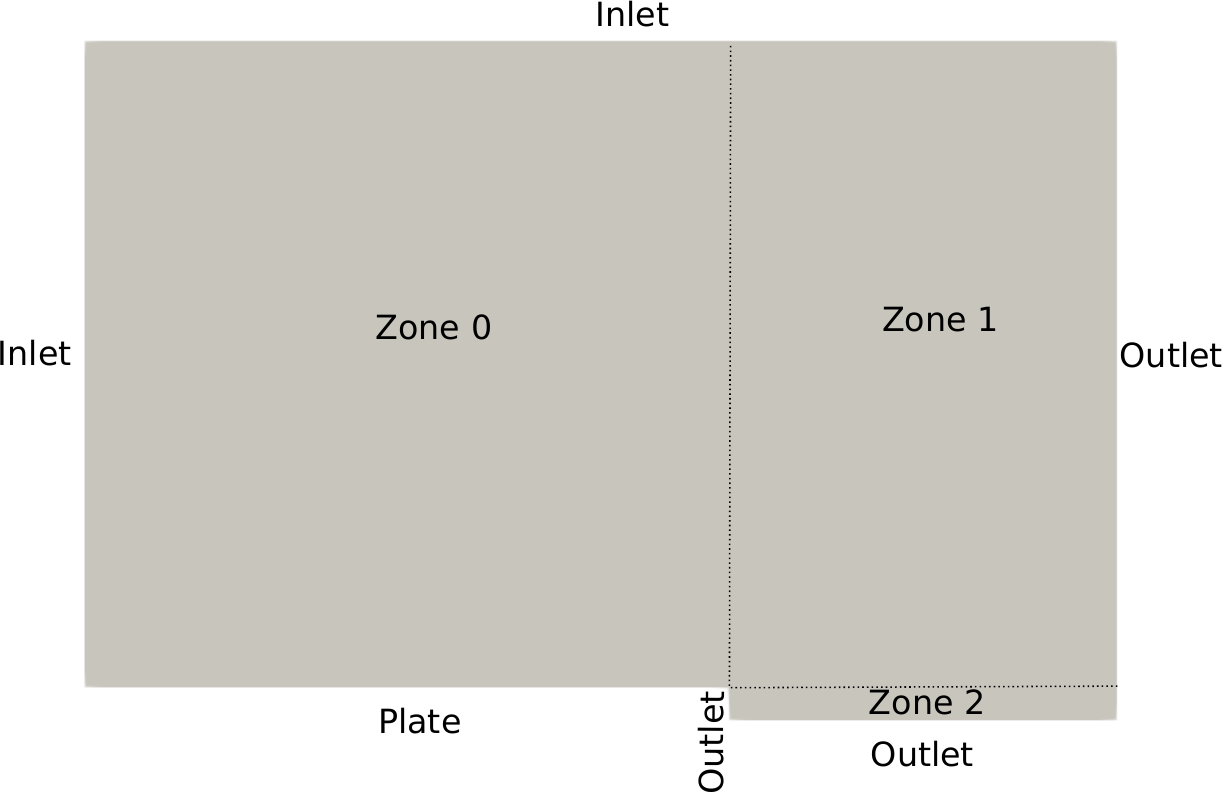}
		\caption{
			\label{fig:meshTopology}  
			Computational domain and boundary conditions.  
		}
	\end{figure}

	This gas-surface interaction problem is derived from the ``hypersonic 3D flow over a flat plate" case, which serves as a benchmarking case for three DSMC codes -dsmcFoam~\cite{white}, DAC and MONACO~\cite{padilla2010comparison}- as well as an experiment~\cite{allegre}. The derived test cases were simulated using dsmcFoam to predict gas-surface interactions, where the flow passes through a control volume, and its bottom patch is defined as a plate, a wall boundary. A new simplified mesh topology based on the ``hypersonic 3D flow over a flat plate" case was created to eliminate flow interactions before the leading edge of the plate by removing block zones; thereby making the inlet and plate patches adjacent. Additionally, some outlet zones were removed to simplify the mesh geometry and reduce the computational cost, as shown in Figure~\ref{fig:meshTopology}. 
		
	An extensive simulation matrix is used to present the effects of flow parameters and wall models on gas-surface interactions. The geometry in Figure~\ref{fig:meshTopology} is used for all cases. The simulation matrix begins with the same initial conditions as in the benchmarked cases~\cite{white,padilla2010comparison,allegre}, varying AoAs and wall models. Subsequently, different flow conditions are created by maintaining the same free-stream pressure of 0.06833 $Pa$ from the benchmark case. The temperature and velocity of the free stream are gradually increased from the values of the benchmark case to higher levels, converging to possible orbital values while observing gas-surface interactions with four different wall models. The AoA values are selected with a larger margin to clearly explore the effects of AoA patterns. The simulation matrix is shown in Table~\ref{table:simulationMatrix}. 
	
	\begin{table*}
		\centering
		\caption{Simulation matrix.} 
		\footnotesize{
			\begin{tabular}{ccllc}
				\hline
				\textbf{Simulation Group}      & \textbf{Species} & \textbf{Macroscopic Values} & \textbf{Wall Model \& AoA} & \textbf{Number of Simulations} \\
				\hline
				\#1               & N$_{2}$ & 13 $K$             & Specular - 0$^{\circ}$, 15$^{\circ}$, 30$^{\circ}$  & 12 \\
				                  &  & 1500 $m/s$           & Maxwellian - 0$^{\circ}$, 15$^{\circ}$, 30$^{\circ}$ & \\				
				                  &  & Kn = 0.0168        & Diffuse - 0$^{\circ}$, 15$^{\circ}$, 30$^{\circ}$ & \\				
				                  &  &                    & CLL - 0$^{\circ}$, 15$^{\circ}$, 30$^{\circ}$ & \\ \addlinespace 
                \#2               & N$_{2}$ & 300 $K$            & Specular - 15$^{\circ}$  & 4\\ 
                			      &    & 1500 $m/s$         & Maxwellian - 15$^{\circ}$ &  \\				
                				  &    & Kn = 0.8         & Diffuse - 15$^{\circ}$ & \\				
                                  &    &                  & CLL - 15$^{\circ}$ & \\ \addlinespace 			
				\#3               & N$_{2}$ & 500 $K$            & Specular - 15$^{\circ}$ & 4 \\
								  &    & 1500 $m/s$         & Maxwellian - 15$^{\circ}$ & \\				
				                  &    & Kn = 1.5         & Diffuse - 15$^{\circ}$ & \\				
				                  &    &                  & CLL - 15$^{\circ}$ & \\ \addlinespace 
				\#4               & N$_{2}$ & 500 $K$            & Specular - 15$^{\circ}$ & 4 \\
				                  &    & 4500 $m/s$         & Maxwellian - 15$^{\circ}$ & \\				
				                  &    & Kn = 1.5         & Diffuse - 15$^{\circ}$ & \\				
				                  &    &                  & CLL - 15$^{\circ}$ & \\ \addlinespace 			
				\#5               & N$_{2}$ & 500 $K$            & Specular - 15$^{\circ}$ & 4 \\
				                  &    & 7500 $m/s$         & Maxwellian - 15$^{\circ}$ & \\				
				                  &    & Kn = 1.5         & Diffuse - 15$^{\circ}$ & \\				
				                  &    &                  & CLL - 15$^{\circ}$ & \\ \addlinespace 
				\#6               & N$_{2}$ & 700 $K$            & Specular - 15$^{\circ}$ & 4 \\
				                  &    & 7500 $m/s$         & Maxwellian - 15$^{\circ}$ & \\				
				                  &    & Kn = 2.3         & Diffuse - 15$^{\circ}$ & \\				
				                  &    &                  & CLL - 15$^{\circ}$ & \\ \addlinespace 			
				\#7               & N$_{2}$ & 700 $K$            & Specular - 15$^{\circ}$ & 4 \\
				                  & O$_{2}$ & 7500 $m/s$         & Maxwellian - 15$^{\circ}$ & \\				
				                  &    & Kn = 2.3         & Diffuse - 15$^{\circ}$ & \\				
				                  &    &                  & CLL - 15$^{\circ}$ & \\ 
				\hline& 
			\end{tabular}
			\label{table:simulationMatrix}
		}
	\end{table*} 

The selected working gases are VHS nitrogen (N$_{2}$) and oxygen (O$_{2}$), with reference diameters of 4.17$\times$ $10^{-10}$ $m$ and 4.07$\times$ $10^{-10}$ $m$, respectively. The viscosity indices, $\omega$, are 0.74 and 0.77, respectively, at a reference temperature of 273 $K$. The translational and rotational energy modes are activated, and the Larsen-Borgnakke method is employed to model rotational energy exchange with a constant relaxation collision number of 5 for VHS~\cite{bird}. 
	
	The plate surface temperature is set at 290 $K$, and four different wall models are employed: fully specular, fully diffuse, Maxwellian mixed diffuse-specular with a diffuse fraction of 0.5, and fully accommodated CLL wall with unity tangential, normal, and rotational accommodation coefficients.
	
	In the initial step of the simulation, DSMC particles are placed at the inlet of the control volume. Before running the DSMC solver, sampling is paused until the simulation reaches a steady state. The simulation is then started. 

	\section{Results} 
	The changes in the number of DSMC particles in the control volume and their average linear kinetic energy were used as indicators to determine whether the solution had reached a steady state. Data were collected from all subsequent time steps, and the global results were presented when the solution reached a steady state. Averaging over a large number of time steps beyond this point helps reduce scatter and statistical error in the global results. The post-processing step was performed using the \textit{sample} tool of OpenFOAM, Paraview 5.11.0, and in-house Python scripts. 
	
\begin{figure} [ht]
		\centering
		\includegraphics[width=0.9\columnwidth]{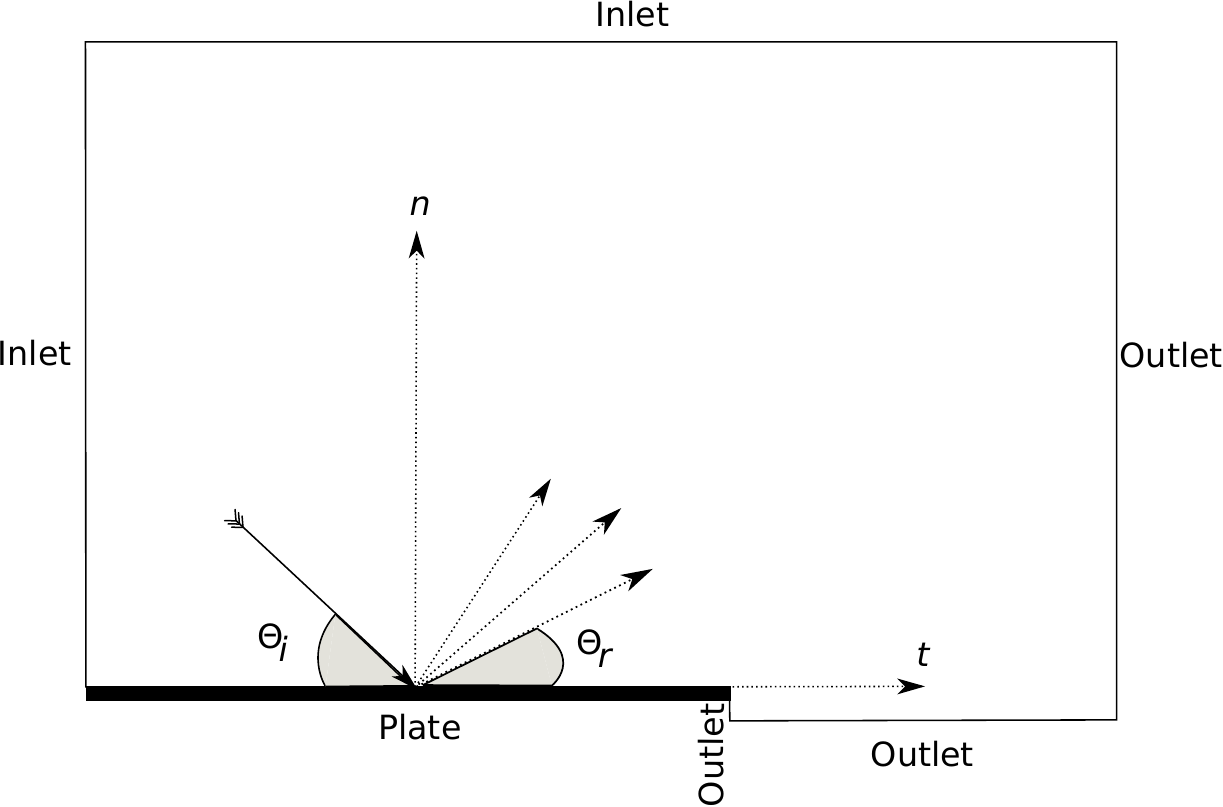}
		\caption{
			\label{fig:particleAngles}  
			Schematic of particle incident and reflection angles. 
		}
	\end{figure}

	The term \textit{force density}, $f_{D}$, in dsmcFoam, differs from that in continuum applications. As stated in Ref.~\cite{white}, \begin{quote} [t]he dsmcVolFields measurement class is designed to detect the surface normal vector of each face on the boundaries and convert the measured force density into the pressure and shear stress components.\end{quote} Thus, the drag force is expressed as $F_{drag} = \int_{A} f_{D} dA$, and the drag coefficient, $Cd$, is calculated as:
	
	\begin{equation}
	\label{cp} 
	C_{d}=\frac{f_{D}}{{\sfrac{1}{2}}\rho_{\infty}U^2_{\infty}},
	\end{equation}

	\noindent
	where $\rho_{\infty}$ is the density of the free stream and $U_{\infty}$ is the velocity of the free stream. The normalised density is $\varphi={n_{surface}}/{n_{\infty}}$, where $n_{surface}$ is the measured number density on the surface at the latest time step of steady state and $n_{\infty}$ is the number density of the free stream flow. The angle of incidence of the gas particles with the surface, $\Theta_{i}$, and the reflection angle, $\Theta_{r}$, are illustrated in Figure~\ref{fig:particleAngles}. 
	
\subsection{AoA} \label{sec:AoA}
	For the investigation of the effect of the AoA, simulation group 1 in Table~\ref{table:simulationMatrix} is used, with the temperature of the N$_{2}$ free stream, $T_{\infty}$, of 13.32 $K$ and the velocity of the free stream, $U_{\infty}$, of 1503 $m/s$. These initial conditions are identical to the benchmark case in dsmcFoam~\cite{white}. The molecular number density, $n_{\infty}$, is calculated to be 3.716 $\times$ $10^{20}$ $1/m^{3}$. Each representative DSMC particle simulates 4.645 $\times$ $10^{9}$ real molecules, and the time step is chosen to be 3.102 $\times$ $10^{-9}$ seconds to accurately capture intermolecular collisions. The total Kn is 0.0168, placing the flow in the slip regime. Results for three different AoAs—0$^{\circ}$ (parallel flow to the surface), 15$^{\circ}$, and 30$^{\circ}$—will be discussed. 
	
	\subsubsection{Scattering} 
\begin{figure}[H]
		\centering
		\begin{subfigure}{0.48\textwidth}
			\includegraphics[width=\textwidth]{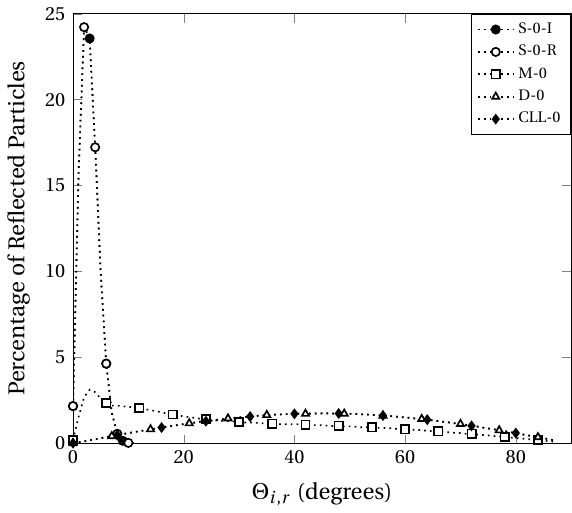}
			\caption{AoA = 0$^{\circ}$.}
			\label{fig:scatteringAoA0}
		\end{subfigure}
		\hfill
		\begin{subfigure}{0.48\textwidth}
			\includegraphics[width=\textwidth]{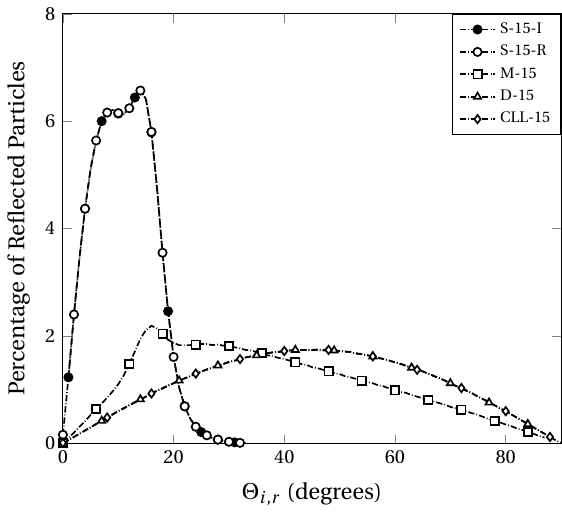}
			\caption{AoA = 15$^{\circ}$.}
			\label{fig:scatteringAoA15}
		\end{subfigure}
		\hfill
		\begin{subfigure}{0.48\textwidth}
			\includegraphics[width=\textwidth]{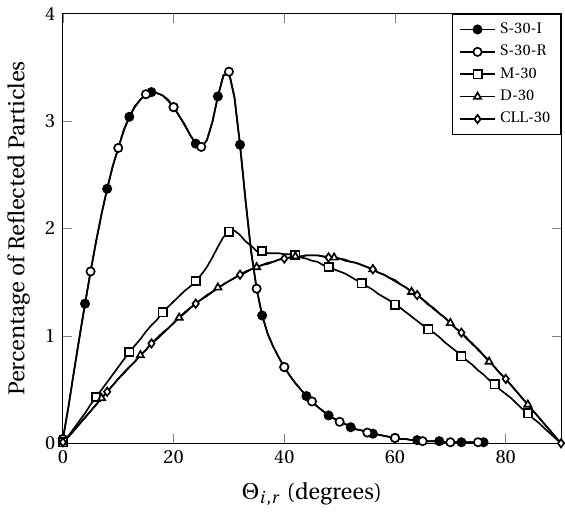}
			\caption{AoA = 30$^{\circ}$.}
			\label{fig:scatterringAoA30}
		\end{subfigure}
		\caption{Interaction angles vs percentage of reflected particles from overall surface of the wall for three different AoAs and four various wall models. In legends, S, M, D, and CLL represent specular, Maxwellian, diffuse, and CLL wall models. 0, 15, 30 represent AoAs, and I is for incident and R is for reflection.} 
		\label{fig:scatteringAoA}
	\end{figure}

	Figure~\ref{fig:scatteringAoA} shows the surface interaction angles of DSMC particles, $\Theta$, for three different AoAs and four wall models. 
	
	The specular wall model shows identical incident and reflection profiles; particles maintain their velocity during the interaction with the plate and change direction while preserving the incident angle as they bounce back from the surface, as expected. Although a pattern closely resembling AoA of the flow is observed, slight variations at the three AoAs are also evident during the interaction with the wall, as seen in Figures~\ref{fig:scatteringAoA0},~\ref{fig:scatteringAoA15}, and~\ref{fig:scatterringAoA30}. 
	
	In Figure~\ref{fig:scatteringAoA0}, for specular reflection, a peak is observed in the reflection angle, with approximately 24.22\% of particles interacting with the surface at 2$^{\circ}$. Near-wall collisions between incoming and reflected molecules can disturb the trajectory of the reflected molecules. Such collisions can result in minor angular deviations.
	
\begin{figure}[H]
		\centering
		\begin{subfigure}{0.5\textwidth}
			\includegraphics[width=\textwidth]{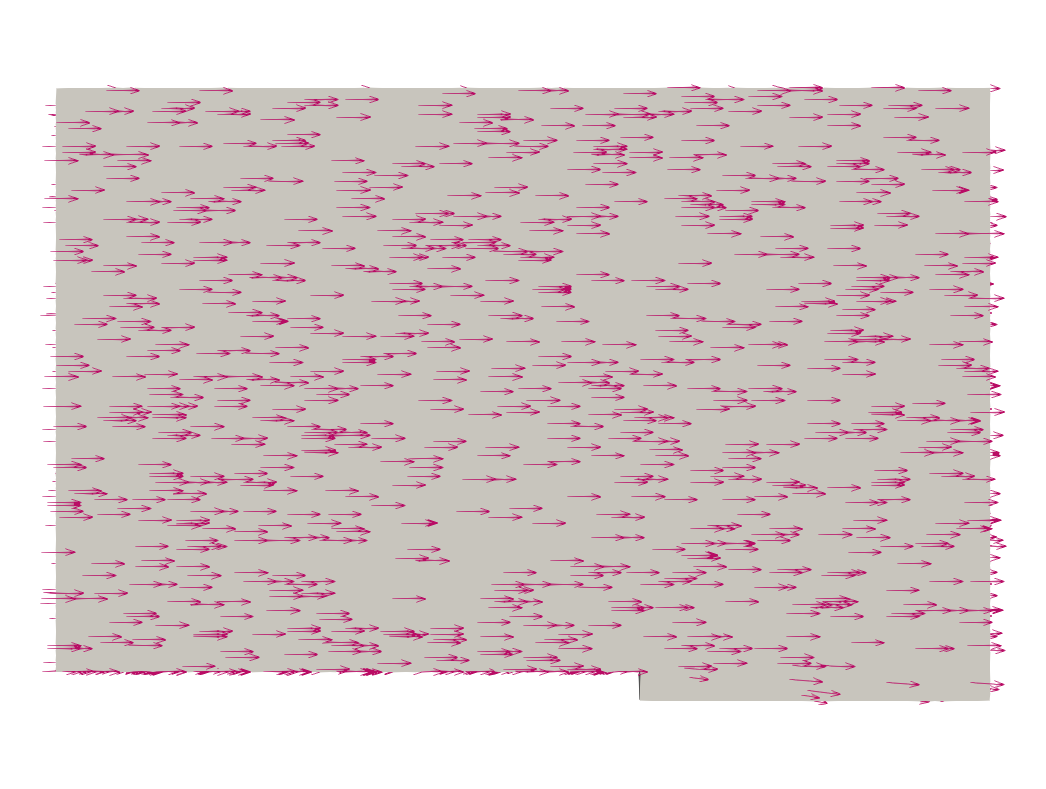}
			\caption{AoA = 0$^{\circ}$, Specular.}
			\label{fig:gylphAoA-S-0}
		\end{subfigure}
		\hfill
		\begin{subfigure}{0.5\textwidth}
			\includegraphics[width=\textwidth]{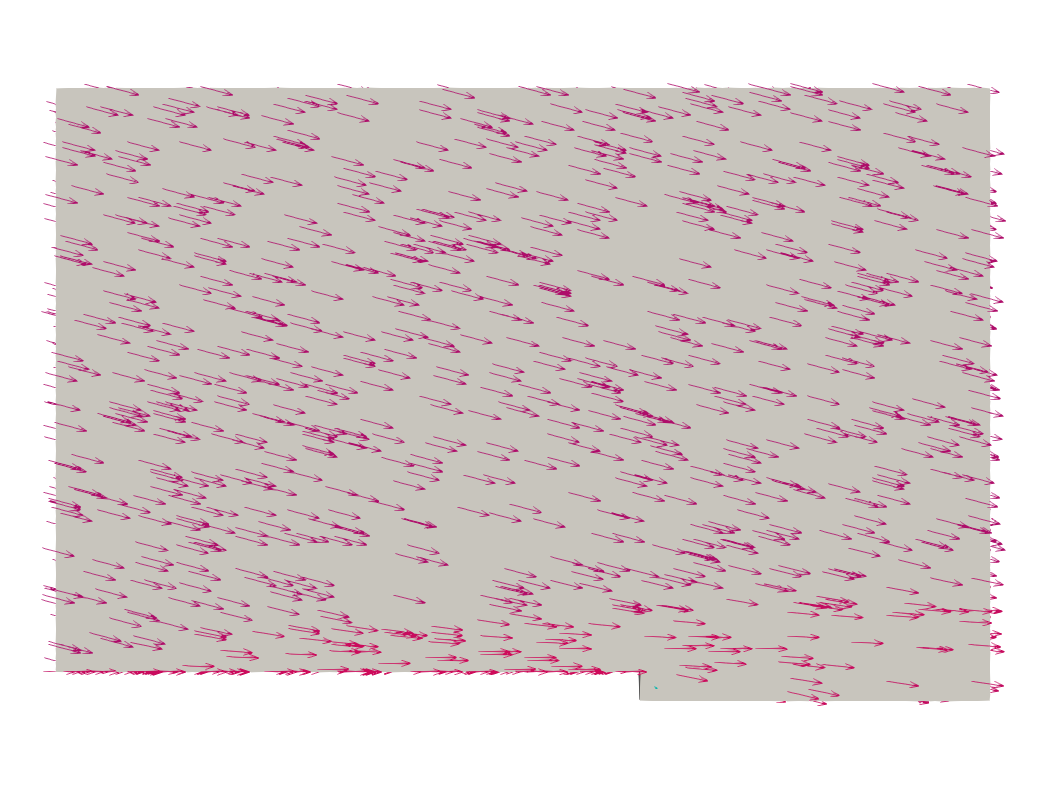}
			\caption{AoA = 15$^{\circ}$, Specular.}
			\label{fig:gylphAoA-S-15}
		\end{subfigure}
		\hfill
		\begin{subfigure}{0.5\textwidth}
			\includegraphics[width=\textwidth]{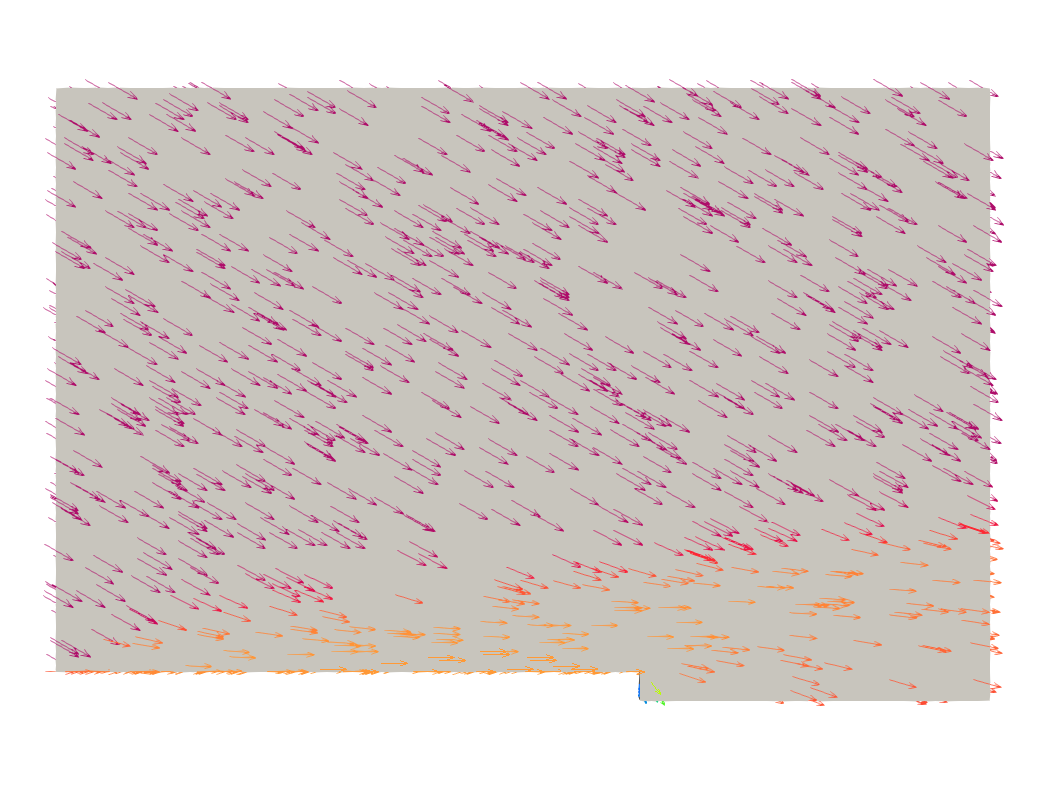}
			\caption{AoA = 30$^{\circ}$, Specular.}
			\label{fig:gylphAoA-S-30}
		\end{subfigure}
		\caption{Flow visualisation of the specular wall model cases for all three AoAs.} 
		\label{fig:gylphAoA-S}
	\end{figure}

	In other specular reflection cases at 15$^{\circ}$ and 30$^{\circ}$ AoAs, the flow strikes the surface at the selected AoA values, creating the highest peak points on the plots, with values of 6.5\% and 3.45\% for 15$^{\circ}$ and 30$^{\circ}$ AoAs, respectively, near the leading edge of the wall, as seen in Figures~\ref{fig:scatteringAoA15} and~\ref{fig:scatterringAoA30}. However, moving further from the leading edge through the trailing edge, the downstream flow in the vicinity of the wall is influenced by the incoming flow, and 6.21\% and 3.21\% of N$_{2}$ molecules interact with the surface at 9$^{\circ}$ and 16$^{\circ}$ AoAs, respectively, for the original 15$^{\circ}$ and 30$^{\circ}$ AoA cases, which are smaller peaks in the plots. Figure~\ref{fig:gylphAoA-S} shows the flow direction in the specular reflection cases at three AoAs.
	
	Figure~\ref{fig:scatteringAoA} also shows the reflection angle versus reflected particle percentage for the Maxwellian wall patch with a diffuse factor of 0.5. The reflection angles exhibit a Maxwellian distribution trend, but small peaks appear in all AoA values. This wall patch provides a more moderate approach between specular and diffuse wall, as expected. 
	
	When Figure~\ref{fig:scatteringAoA} is examined for fully diffuse and fully accommodated walls —diffuse and CLL wall models— the particle reflection behaviours are predicted exactly the same for all three AoAs. In both models and for all AoAs, a Maxwellian distribution pattern is observed for the particle reflections, as expected, with a maximum percentage of reflected particles of 1.75\% between 40$^{\circ}$  and 50$^{\circ}$. 
	
\begin{figure}[H]
		\centering
		\begin{subfigure}{0.5\textwidth}
			\includegraphics[width=\textwidth]{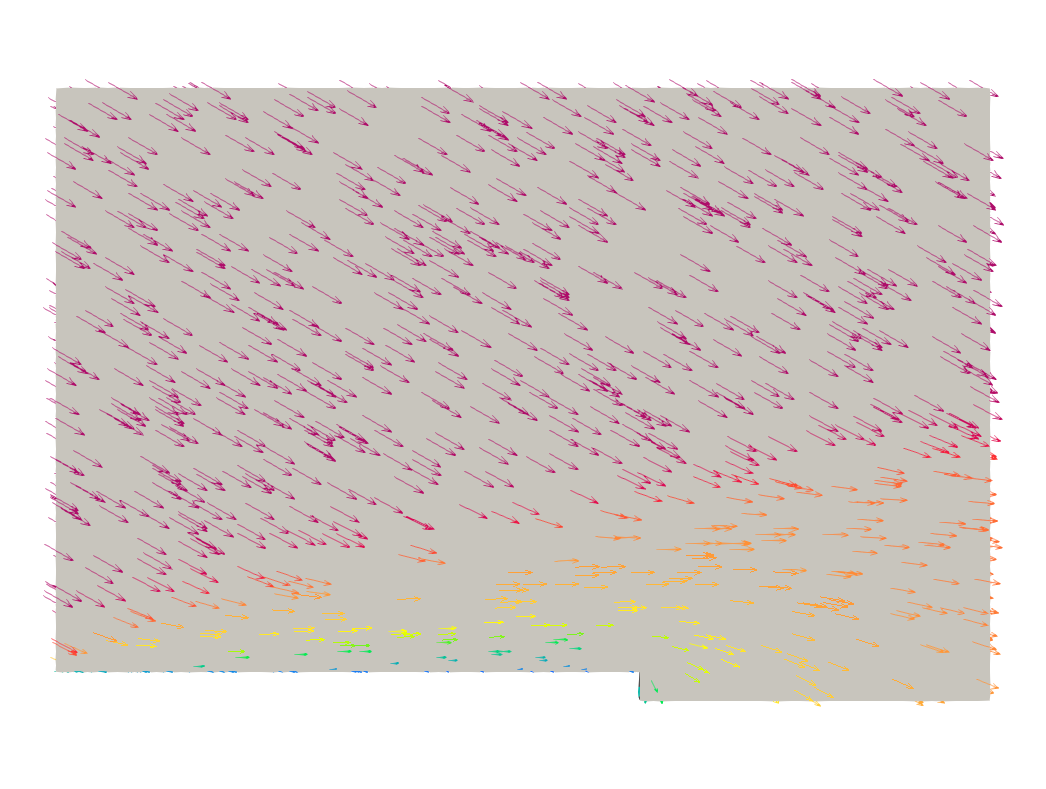}
			\caption{AoA = 30$^{\circ}$, Maxwellian.}
			\label{fig:gylphAoA-M-30}
		\end{subfigure}
		\hfill
		\begin{subfigure}{0.5\textwidth}
			\includegraphics[width=\textwidth]{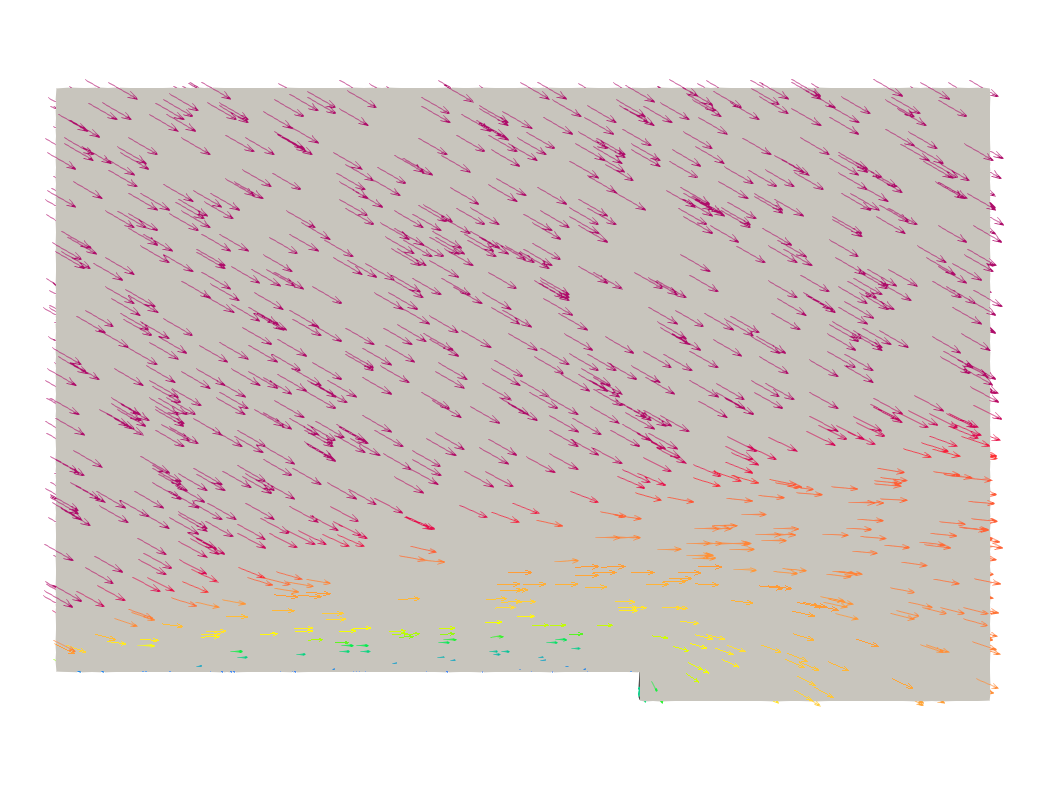}
			\caption{AoA = 30$^{\circ}$, Diffuse.}
			\label{fig:gylphAoA-D-30}
		\end{subfigure}
		\hfill
		\begin{subfigure}{0.5\textwidth}
			\includegraphics[width=\textwidth]{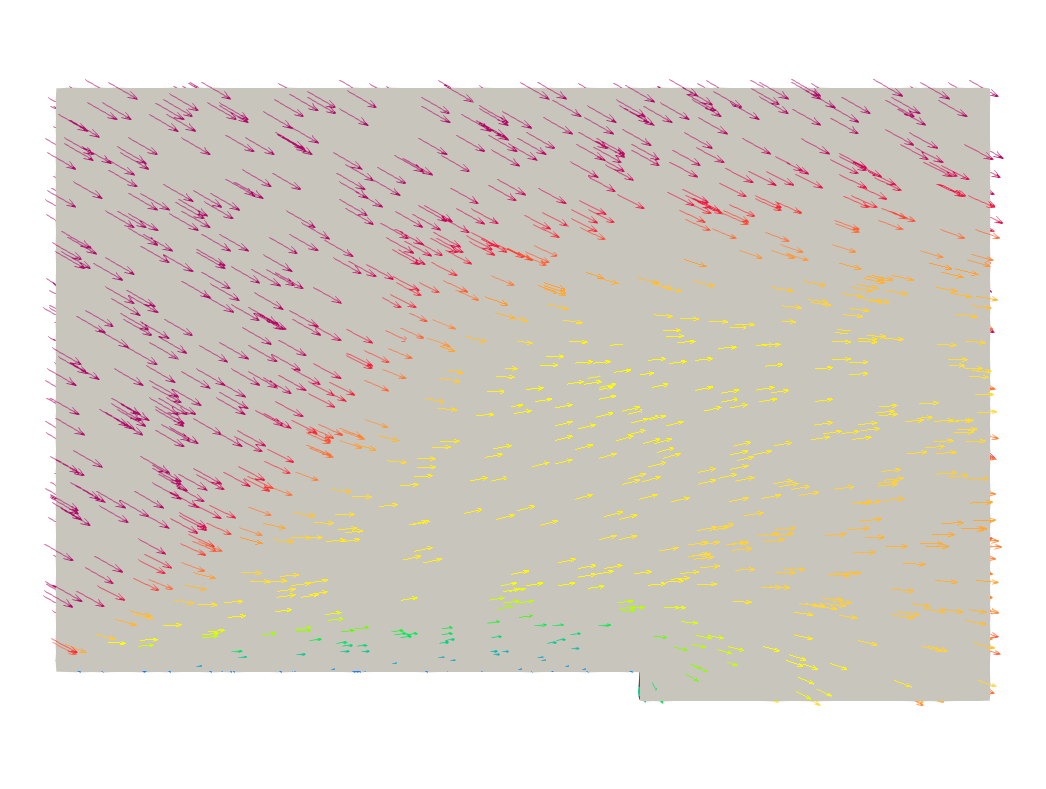}
			\caption{AoA = 30$^{\circ}$, CLL.}
			\label{fig:gylphAoA-CLL-30}
		\end{subfigure}
		\caption{Flow visualisation of the Maxwellian, diffuse, and CLL wall model cases for 30$^{\circ}$ AoA.} 
		\label{fig:gylphAoA-otherWallModels30}
	\end{figure}

	A filter through the control volume is employed to visualise the flow patterns for specular cases at all AoAs, as well as for the other three wall models at an AoA of 30$^{\circ}$ using the velocity vectors in Figures~\ref{fig:gylphAoA-S} and~\ref{fig:gylphAoA-otherWallModels30}, respectively. To clearly illustrate the changes in the flow direction, which are coloured using orientation arrays as the velocity vectors of the flow. In Figure~\ref{fig:gylphAoA-S}, the effect of AoA on the flow in the vicinity of the wall is shown using the specular wall model. Additionally, the impact of the wall model is demonstrated in Figures~\ref{fig:gylphAoA-S-30} and~\ref{fig:gylphAoA-otherWallModels30} at 30$^{\circ}$ AoA. It is observed that the flow is more disturbed in the higher AoA. The CLL model also creates a more disturbed flow in the control volume compared to other wall models. 
	
	Alterations in intermolecular and particle-surface interactions are observed when the wall model is varied. Furthermore, at higher AoAs, the intermolecular and gas-surface interactions increase. 

\pagebreak 

	\newpage
	
	\pagebreak

	\subsubsection{Surface parameters}
\begin{figure} [h!]
		\centering
		\includegraphics[width=0.6\columnwidth]{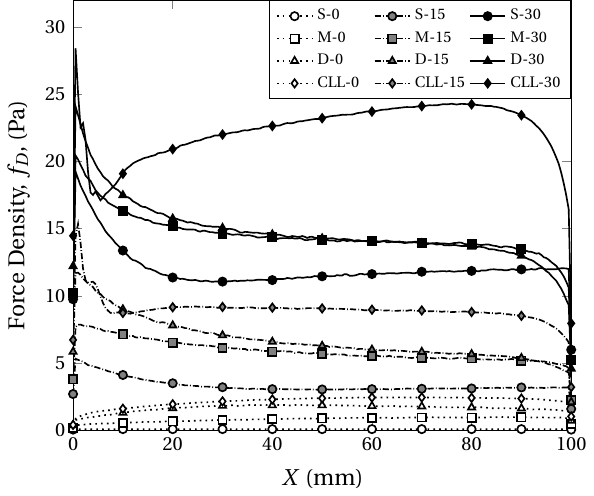}
		\caption{
			\label{fig:fdAoA}  
			The effect of AoA and wall model on the force density. 
		}
	\end{figure}

	Figure~\ref{fig:fdAoA} shows the variation in the force density as a function of AoA and wall model. It is observed that the CLL wall model predicts the highest surface force for each AoA group, followed by the diffuse, Maxwellian, and specular wall models. Furthermore, an increase in AoA results in higher surface forces. 
    
\begin{figure} [h!]
		\centering
		\includegraphics[width=0.6\columnwidth]{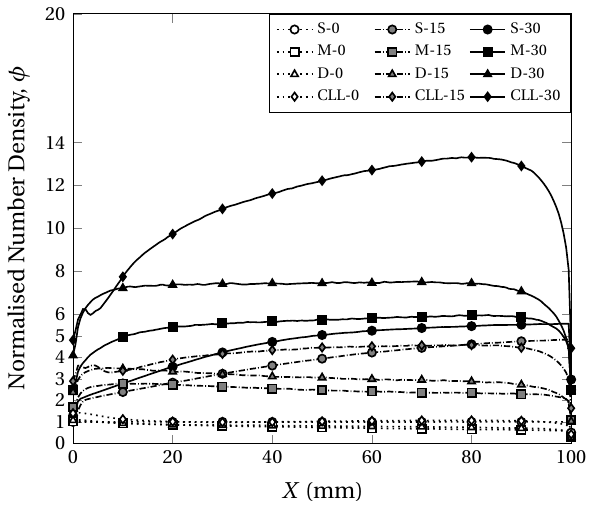}
		\caption{
			\label{fig:numberDensityAoA}  
			The effect of AoA and wall model on the normalised number density.   
		}
	\end{figure}

	The normalised number density on the surface from the leading edge to the trailing edge of the wall for all AoAs and wall models is also shown in Figure~\ref{fig:numberDensityAoA}. The patterns of surface forces and change in the number density can also be associated, which can be deduced in the comparison of Figures~\ref{fig:fdAoA} and~\ref{fig:numberDensityAoA}.  
	
\begin{figure} [h!]
		\centering
		\includegraphics[width=0.7925\columnwidth]{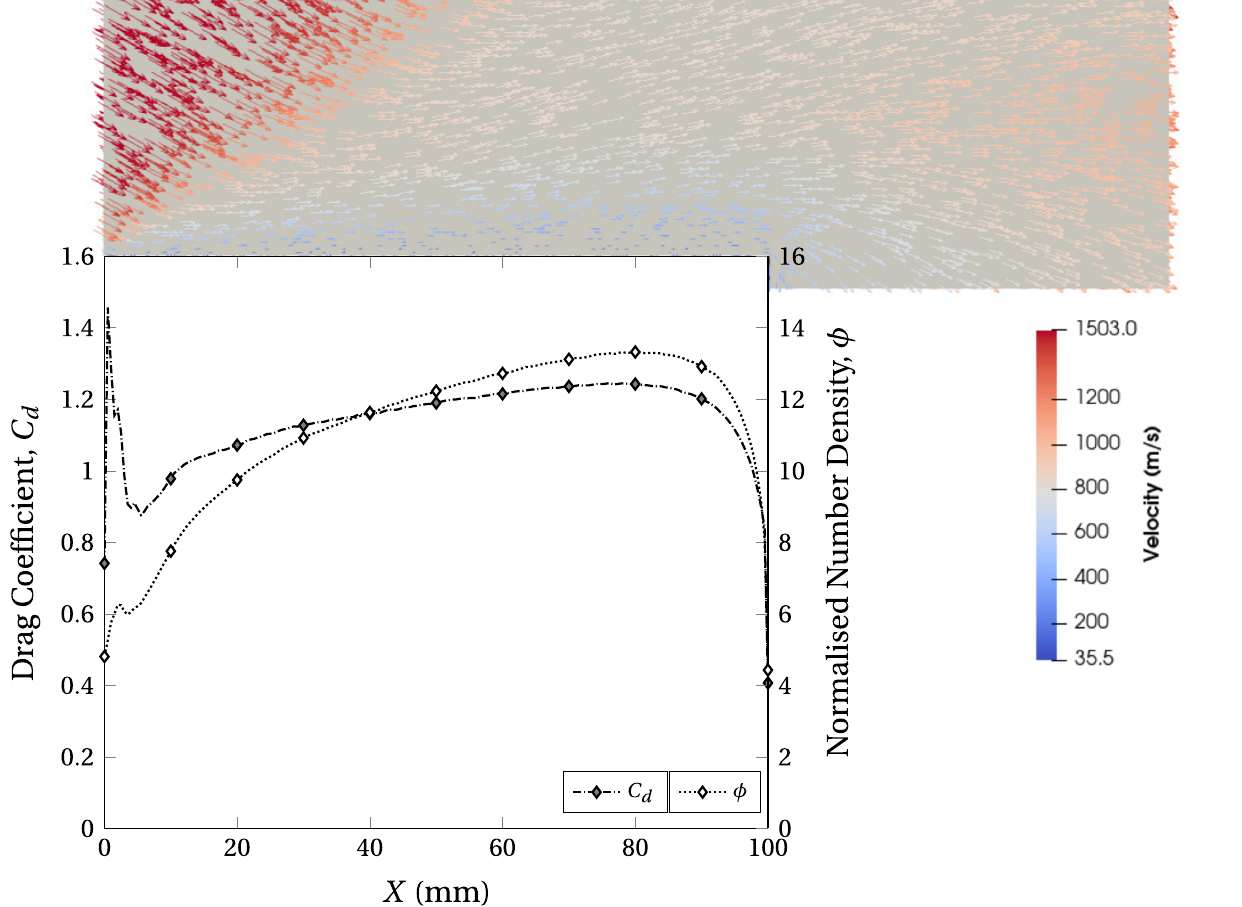}
		\caption{
			\label{fig:30_edittedAoA}  
			Visualisation filter to show flow direction with velocity colouring, and drag and normalised number density values in the vicinity of the surface for the CLL wall model at 30$^{\circ}$ AoA.    
		}
	\end{figure}

	It should be noted that the specular wall model in dsmcFoam is not capable of performing energy updates. The Maxwellian and diffuse wall models randomise scattering by deleting the pre-collision history of particles and complete the energy update using the equipartition theorem with the surface temperature. Physically, in non-parallel flows, it is expected that the flow first interacts with the surface at the leading edge with high velocity and low number density on the surface, then reflecting back. In this case where two inlet patches are defined from left and top of the control volume, this process occurs repeatedly along the wall. The reflecting and incoming number of particles accumulates toward the trailing edge, resulting in a higher number density and surface force at the trailing edge, as shown in Figure~\ref{fig:30_edittedAoA}.
	
	Thus, in Figure~\ref{fig:numberDensityAoA}, it is observed that the specular wall model presents an increasing trend in the number density toward the trailing edge of the wall. However, if the material is not perfectly smooth and rigid, this behaviour should also be modelled using the diffuse fraction or accommodation coefficients. The diffuse wall model simulates the surface forces almost similar to the CLL model around the leading edge -as both wall models are fully accommodated surfaces-, as shown in Figure~\ref{fig:fdAoA}. However, moving toward the trailing edge, the number density is obtained as constant in Maxwellian and diffuse wall models, which does not represent the cumulative interaction through the trailing edge discussed earlier. The flow velocity in the vicinity of the wall is strongly linked to other parameters, as shown in Figure~\ref{fig:30_edittedAoA}.
	
\pagebreak 

	\subsubsection{Velocity profile}
\begin{figure} [h!]
		\centering
		\includegraphics[width=0.6\columnwidth]{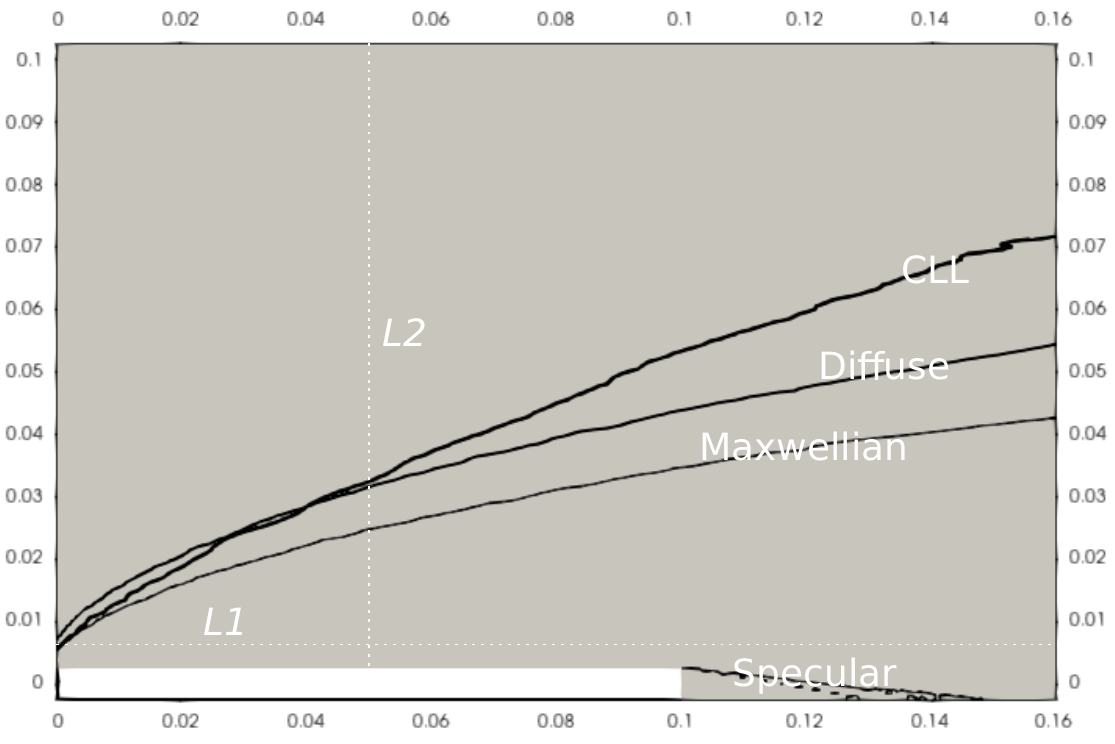}
		\caption{
			\label{fig:U-0-lines}  
			Black lines shows the locations of velocity contours at 1450 $m/s$ for wall models. $L1$ and $L2$ represent the velocity data collection lines.  
		}
	\end{figure}

	As previously discussed, the reflection kernel affects the flow parameters around the surface, creating alterations in flow patterns with the influence of altering AoA. When the velocity contours for the four wall models are plotted together, this effect is clearly observed. The free-stream flow enters the control volume with a velocity of 1503 $m/s$; however, the flow is disturbed around the plate and slows down due to intermolecular and gas-surface interactions. To approximate the location where the velocity deviates from the free-stream value within the control volume, velocity contours for 1450 $m/s$ are plotted for the four wall models at 0$^{\circ}$ AoA, as shown in Figure~\ref{fig:U-0-lines}. The flow velocity of 1450 $m/s$ is chosen as an indicator for comparison of flow disturbance starting locations, representing near-free stream conditions with a 4\% variance. Each black line is labelled with the corresponding wall model type, marking the region where the flow velocity drops to 1450 $m/s$. The region above these lines represents the free-stream velocity for each wall model. As the effects of the reflection kernel and AoA are shown in Figures~\ref{fig:gylphAoA-S} and~\ref{fig:gylphAoA-otherWallModels30}, the velocity contours in Figure~\ref{fig:U-0-lines} can be linked with them. For instance, in the specular wall model, no energy update is performed, and the magnitude of the velocity is not recalculated, allowing the flow to pass over the surface with less disturbance compared to other active wall models. The wall models that incorporate reflection kernels with energy updates are categorised as `active walls' for easier comparison. 
	
	To investigate flow velocity variations, velocity data along $L1$ and $L2$ are extracted. $L1$ is a horizontal line located just above the surface at $Y$ = 0.005 mm, extending from the inlet patch to the outlet patch. The vertical data collection line, $L2$, starts from the midpoint of the surface at $X$ = 0.05 and reaches the top of the control volume, as shown in Figure~\ref{fig:U-0-lines}. 
	
\begin{figure}[h!]
		\centering
		\begin{subfigure}{0.65\textwidth}
			\includegraphics[width=\textwidth]{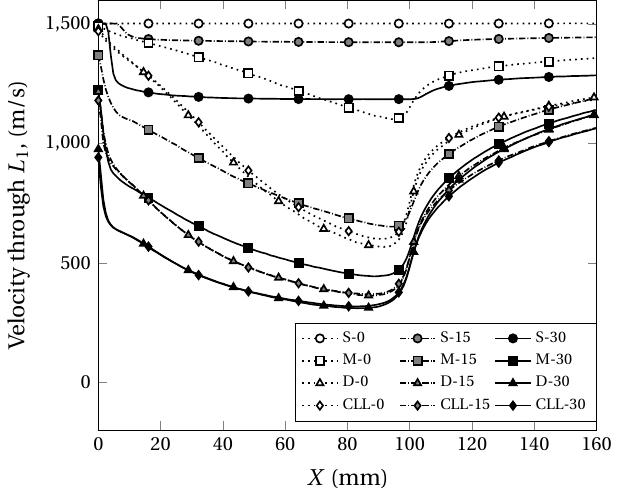}
			\caption{Velocity patterns along the horizontal data collection line.}
			\label{fig:AoA-L1Velocity}
		\end{subfigure}
		\hfill
		\begin{subfigure}{0.65\textwidth}
			\includegraphics[width=\textwidth]{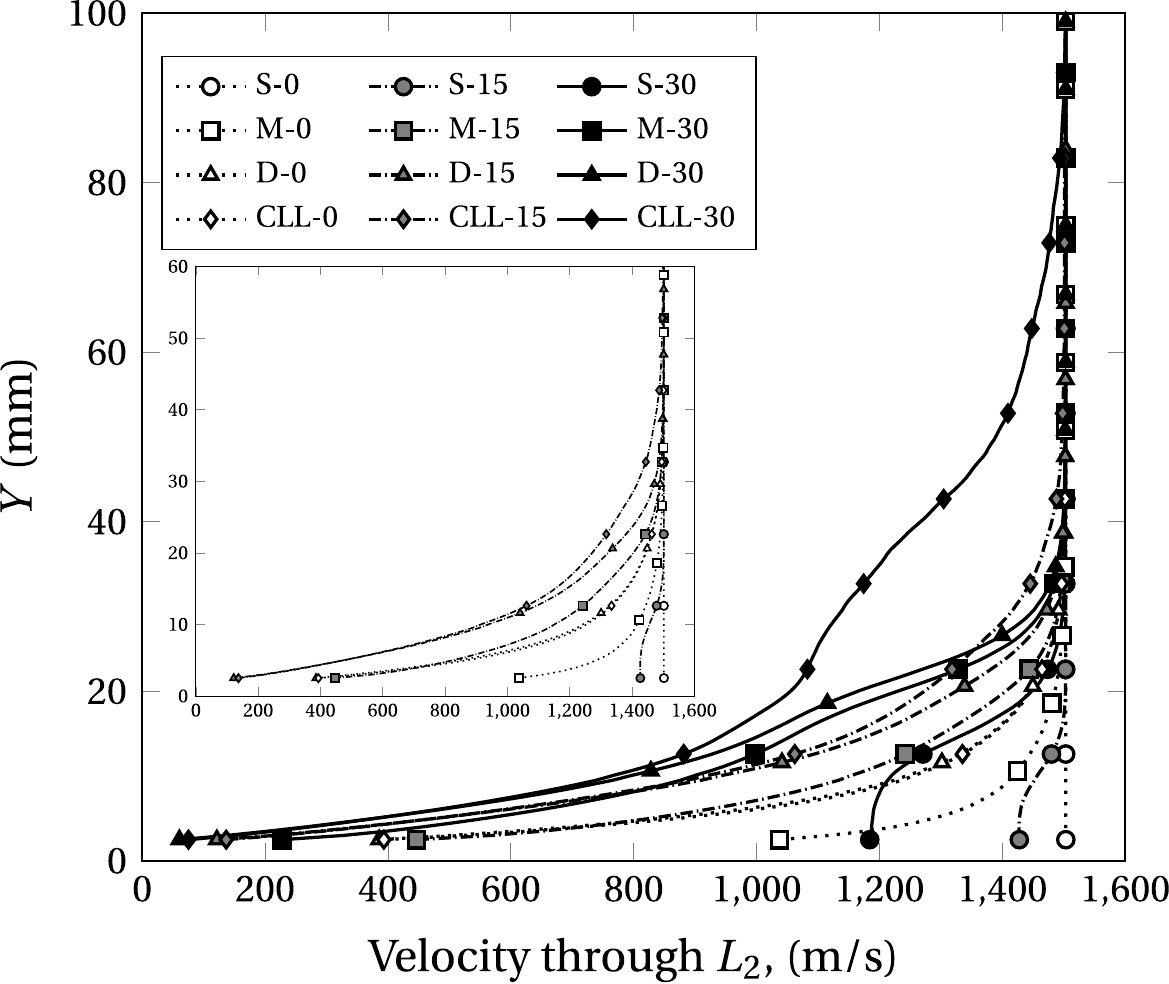}
			\caption{Velocity patterns along the vertical data collection line.}
			\label{fig:AoA-L2Velocity}
		\end{subfigure}
		\caption{Velocity profiles of the specular, Maxwellian, diffuse, and CLL wall models cases at 0$^{\circ}$, 15$^{\circ}$, and 30$^{\circ}$ AoA along the $L1$ and $L2$ data collection lines. The wall models are represented by their initials in the legend, along with the corresponding AoAs. The inset plot in the figure excludes the results for a 30$^{\circ}$ AoA to enhance readability.} 
		\label{fig:AoA-Velocity}
	\end{figure}

	Figure~\ref{fig:AoA-Velocity} shows the change in velocity at the selected locations, through $L1$ and $L2$, for three AoA values and four wall models. As seen in Figure~\ref{fig:AoA-L1Velocity}, the flow is not significantly disturbed in the case of the specular wall at 0$^{\circ}$ AoA. When the AoA is increased to 15$^{\circ}$ and 30$^{\circ}$, the flow gradually slows down, reaches its minimum value, maintains this minimum velocity across the surface, and then leaves the surface with a slightly higher velocity. In contrast, the flow velocity consistently decreases along the surface in the other three wall models. The diffuse and CLL wall models exhibit similar velocity magnitudes. Although the post-collision parameters in the diffuse and CLL wall models, which are fully accommodated, are calculated using different approaches, the post-collision velocity is primarily determined as the most probable velocity modified by the effect of diffuse/accommodation coefficients on the velocity components. The most probable velocity is $\sqrt{2kT/m}$, where $k$ is the Boltzmann constant, $T$ is the surface temperature (290 $K$ for all cases), and $m$ is the molecular mass. Both the diffuse fraction for the diffuse wall and the accommodation coefficients for the CLL wall are unity. Consequently, the velocity patterns of the Maxwellian wall model, with a diffuse fraction value of 0.5, falls between the specular wall and the two fully active walls in terms of velocity behaviour. For all wall models, higher AoA reduces the velocity around the surface. 
	
	Figure~\ref{fig:AoA-L2Velocity} illustrates the change in flow velocity along the data collection line $L2$. In the specular wall model case at 0$^{\circ}$ AoA, disturbance in the flow is not observed from the surface to the top of the control volume. However, the effect of AoA is evident in the specular cases, where the velocity drops to approximately 1400 $m/s$ and 1200 $m/s$ at 15$^{\circ}$ and 30$^{\circ}$ AoA, respectively, around the surface. A gradual decrease in flow velocity near the surface is observed with increasing AoA in all other wall models. The diffuse and CLL wall models exhibit a similar trend, while the Maxwellian wall model lies between the specular and active wall models. However, beyond a height of 20 mm within the control volume, the velocity profiles and magnitudes of the diffuse and CLL wall models deviate, with the CLL wall model showing more disturbed flow at 30$^{\circ}$ AoA. 

\pagebreak 

\subsection{Free-stream Temperature} \label{sec:temperature} 
	In this section, simulation groups 1, 2, and 3 are compared with free-stream temperatures, $T_{\infty}$, of 13.32 $K$, 300 $K$, and 500 $K$, respectively. The molecular number densities, $n_{\infty}$, are calculated as 3.716 $\times$ $10^{20}$ $1/m^{3}$, 1.65 $\times$ $10^{19}$ $1/m^{3}$, and 9.902 $\times$ $10^{18}$ $1/m^{3}$, resulting in each representative DSMC particle simulating 4.645 $\times$ $10^{9}$, 2.062 $\times$ $10^{8}$, 1.237 $\times$ $10^{8}$  real molecules for simulation groups 1, 2, and 3, respectively. The total Kn are 0.0168, 0.8, and 1.5 for simulation groups 1, 2, and 3, respectively, indicating a shift in the flow regime from the slip regime to the transition regime. The free-stream velocity, $U_{\infty}$, and AoA are maintained at 1503 $m/s$, and 15$^{\circ}$, respectively. 
	
	\subsubsection{Scattering}
\begin{figure}[h!]
		\centering
		\includegraphics[width=0.6\columnwidth]{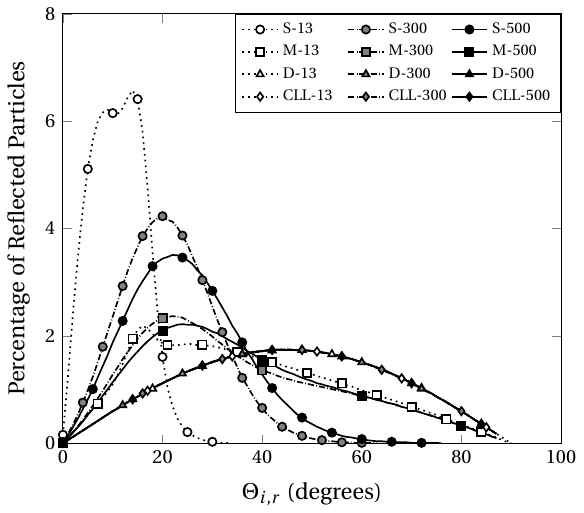}
		\caption{Interaction angles vs percentage of reflected particles from overall surface of the wall for three different free-stream temperatures and four various wall models. S, M, D, and CLL represent specular, Maxwellian, diffuse, and CLL wall models. 13, 300, 500 represent free-stream temperatures.} 
		\label{fig:scatteringTemperature}
	\end{figure}

\begin{figure}[h!]
		\centering
		\begin{subfigure}{0.5\textwidth}
			\includegraphics[width=\textwidth]{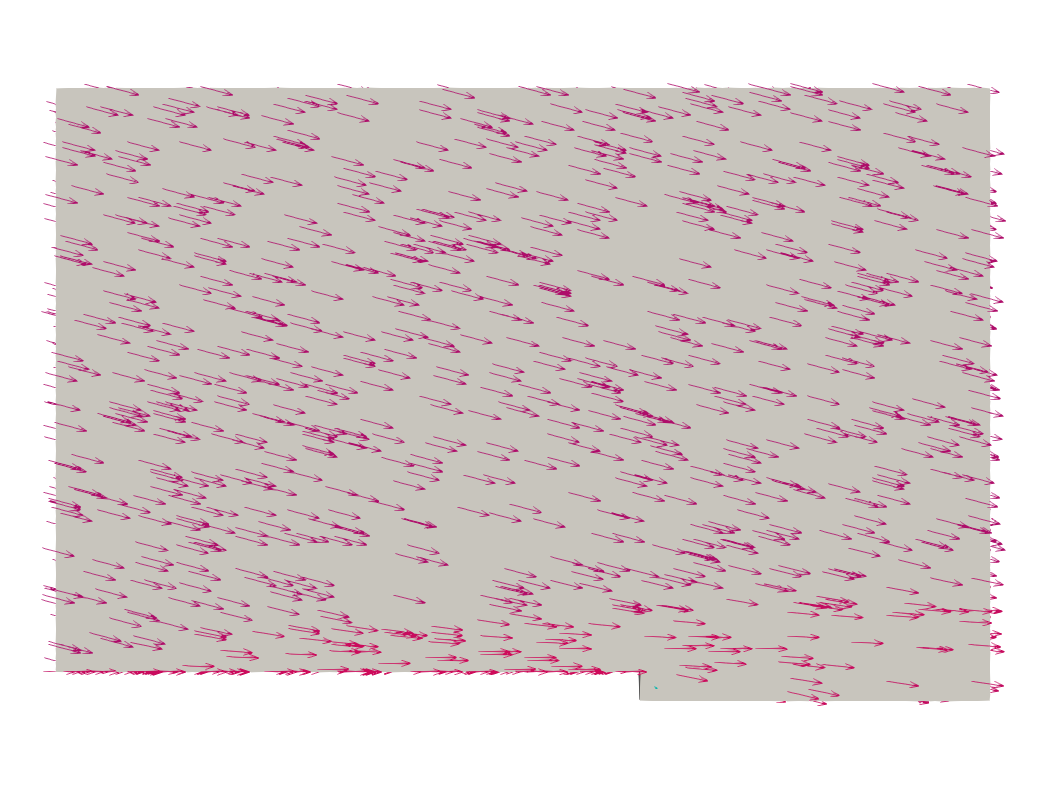}
			\caption{$T_{\infty}$ = 13.32 $K$, Specular.}
			\label{fig:gylphTemp-S-0}
		\end{subfigure}
		\hfill
		\begin{subfigure}{0.5\textwidth}
			\includegraphics[width=\textwidth]{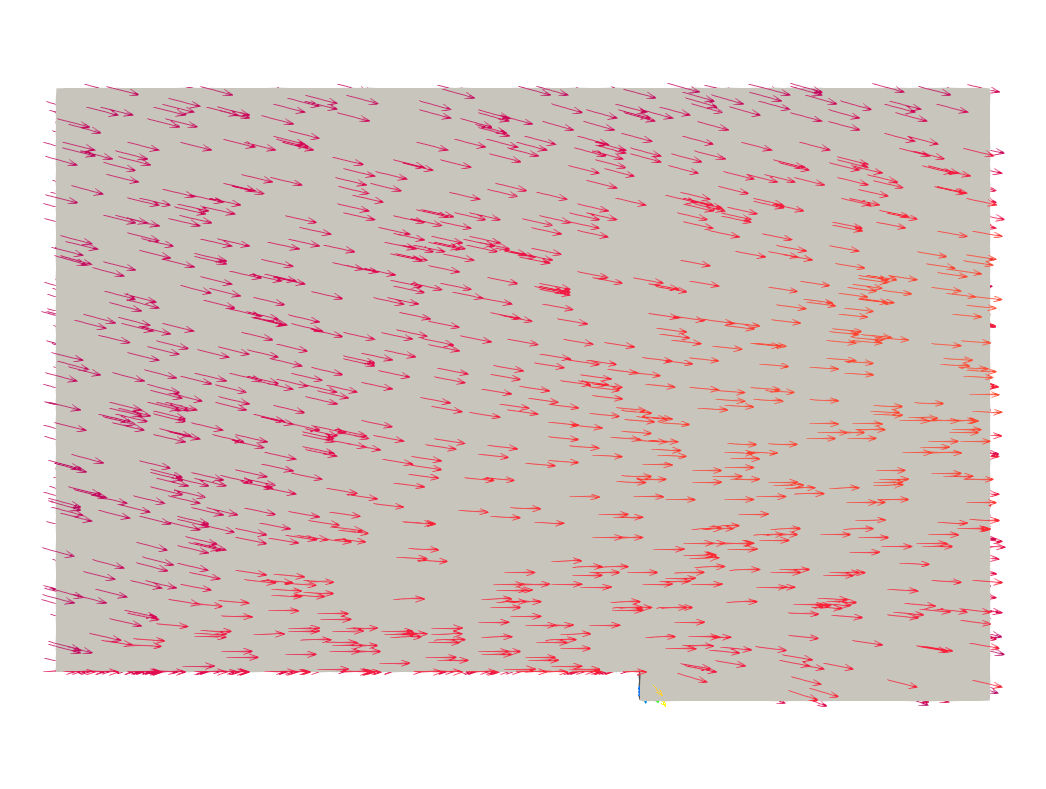}
			\caption{$T_{\infty}$ = 300 $K$, Specular.}
			\label{fig:gylphTemp-S-15}
		\end{subfigure}
		\hfill
		\begin{subfigure}{0.5\textwidth}
			\includegraphics[width=\textwidth]{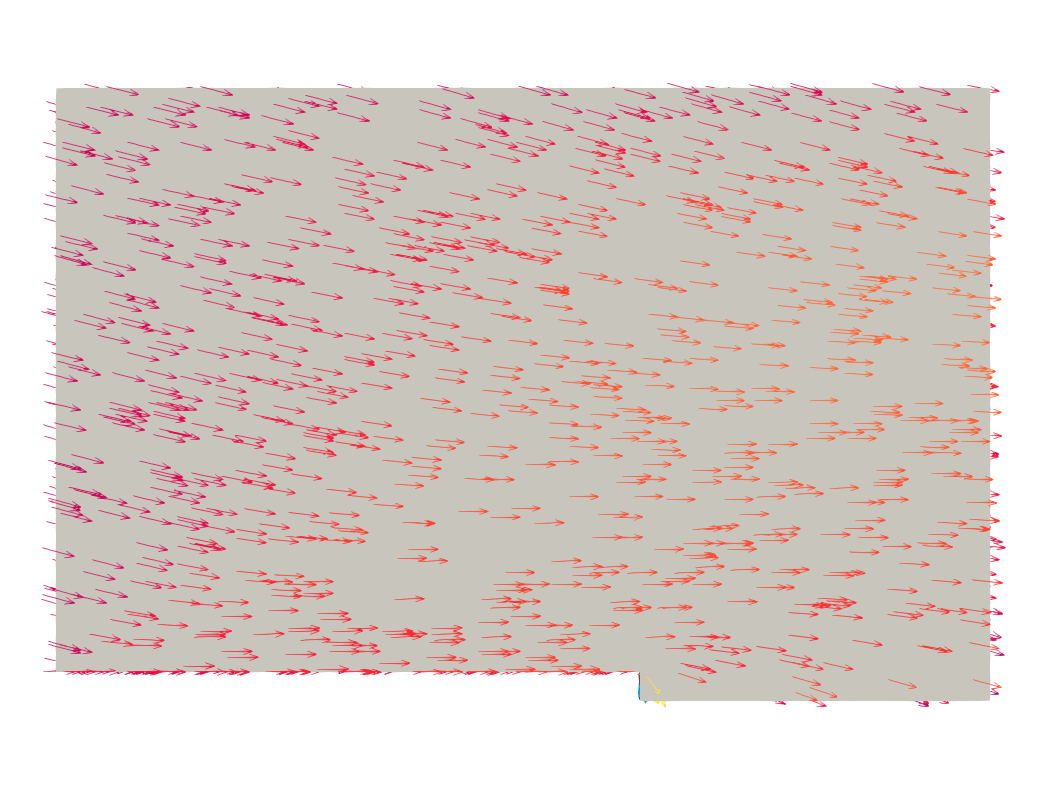}
			\caption{$T_{\infty}$ = 500 $K$, Specular.}
			\label{fig:gylphTemp-S-30}
		\end{subfigure}
		\caption{Flow visualisation of the specular wall model cases for all three free-stream temperatures.} 
		\label{fig:gylphTemp-S}
	\end{figure}

	Figure~\ref{fig:scatteringTemperature} shows the reflection trends of DSMC particles at varying rarefaction levels. As previously mentioned in \S~\ref{sec:AoA}, for the specular wall, the higher peak indicates that 6.5\% of particles reflect back with the surface at 15 degrees, while a smaller peak also occurs where 6.21\% of particles reflect back with the surface at 9 degrees. By increasing Kn, a low-density and expanded flow region forms, and the strength of the flow reduces~\cite{agir2022effect}. The flow is visualised using the flow direction visualisation filter, as shown in Figure~\ref{fig:gylphTemp-S}, where the flow disturbance increases with an increment in Kn. Consequently, most interactions occur slightly higher than the targeted AoA, with 4.23\% and 3.51\% of particles reflecting the surface at 20$^{\circ}$ and 22$^{\circ}$, respectively, at 300 K and 500 K in specular wall model simulations. Similarly, 2.36\% and 2.22\% of particles reflect from the surface at 22$^{\circ}$ and between 23$^{\circ}$–26$^{\circ}$, respectively, at 300 $K$ and 500 $K$ in the Maxwellian wall model. However, as reflections are fully randomised in the diffuse and CLL wall models, the particle reflections maintain the same trend at 300 $K$ and 500 $K$ as observed in the cold flow at 13.32 $K$. 
	
	The gas-surface interactions are also influenced by the temperature of the particles. When the gas temperature increases, their kinetic energy rises, leading to more frequent and energetic collisions with the surface, which can enhance the gas-surface interactions. Additionally, the nature of the gas-surface interaction is often characterised by accommodation coefficients, which describe how fully gas molecules equilibrate with the surface upon collision. For instance, higher accommodation coefficients in the CLL wall model indicate a greater degree of interaction between the gas particles and the surface. Therefore, even the surface temperature is held constant and the gas particle temperature is increased, the increased kinetic energy of the particles can lead to more significant interactions with the surface. However, the extent of this interaction also depends on the accommodation coefficients characterising the gas-surface interaction. Surfaces with higher accommodation coefficients will exhibit more substantial interactions with gas particles, leading to greater energy and momentum exchange. 
	
\pagebreak 
    
	\subsubsection{Surface parameters}
\begin{figure} [ht]
		\centering
		\includegraphics[width=0.6\columnwidth]{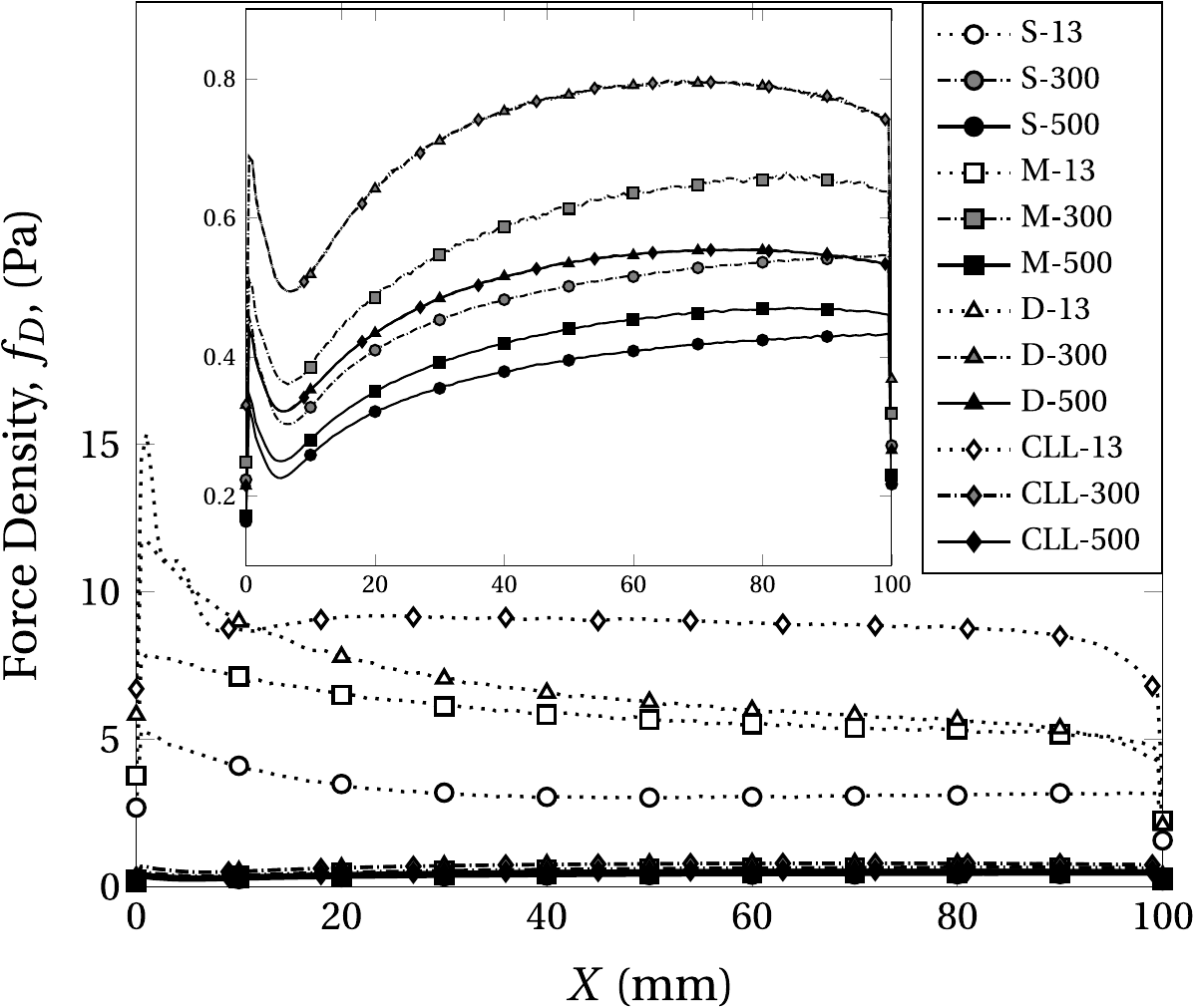}
		\caption{
			\label{fig:fdTemperature}  
			The effect of free-stream temperature and wall model on the force density. The inset plot in the figure provides a close-up view of the results for the cases at 300 $K$ and 500 $K$.    
		}
	\end{figure}

	Figure~\ref{fig:fdTemperature} shows the variations in the force density on the surface due to the free-stream temperature and wall model. The surface forces increases with the increment in diffuse fraction or accommodation coefficient. However, it is observed that the cold free-stream case, at 13.32 $K$, interaction with a warmer wall, at 290 $K$, creates higher surface forces, comparing to warmer flows. Additionally, while the CLL wall model predicts the highest surface forces, the diffuse wall model diverts from the CLL results around $X$ = 10 mm horizontal distance on the surface in the cold flow case. At higher free-stream temperatures, 300 $K$ and 500 $K$, which are greater than the surface temperature of 290 $K$, the surface forces reduce, and the fully active walls, the CLL and diffuse wall models, produce similar results. 
	
	In the cold flow, under the diffuse boundary condition, particles are fully randomised upon reflection, and their post-collision velocities are determined solely by the wall temperature. Since the wall temperature is much higher than the flow temperature, the reflected particles gain significant energy. As a result, the momentum transfer to the wall is less because the reflected particles carry more momentum away from the surface. The CLL model retains the pre-collision memory of particles. This means the post-collision velocity in the CLL wall model is influenced by both the pre-collision velocity and the wall temperature. This leads to higher momentum transfer to the wall compared to the diffuse wall model, and, consequently, a higher surface force for the CLL model. 
	
	When the flow temperature is increased, the thermal energy of the particles enhances. Both the diffuse and CLL boundary conditions fully thermalise the reflection velocities of particles by the wall temperature when the diffuse fraction and accommodation coefficients are unity. Therefore, in the cold flow, the wall temperature dominates the post-collision behaviour, creating significant differences in momentum transfer. Conversely, in the high-temperature flow, highly energetic particles reduce the effect of accommodation treatments, leading to similar results for the diffuse and CLL wall models. 
	
	For the specular wall, the energy or velocity randomisation does not occur, and the thermalisation by the wall temperature does not take place. However, surface forces reduces with the increment in the flow temperature. The thermal velocities of particles are scaled as $\sqrt{T_{particle}}$ according to the Maxwell-Boltzmann distribution. Therefore, increase in the flow temperature leads to high energy and thermal velocity particles in a less dense environment. 
	
	The Maxwellian wall model represents a behaviour between the specular and diffuse wall models, but slightly closer to the diffuse wall model. The diffuse fraction of 0.5 is chosen for the Maxwellian wall model, meaning a significant proportion of particles are treated as diffusely reflected. Half of the particles undergo velocity randomisation upon reflection, which dominates the behaviour compared to the specular reflections of the remaining half of the particles. 
	
	The change in the number density on the surface can be seen due to the dynamics of gas-surface interactions and the flow rarefaction level. As previously discussed, when the particles reflect off the surface, they create a localised increase in the number density. Additionally, as the flow continues along the surface, the cumulative effect of reflected particles builds up, leading to a gradual increase in number density from leading edge to the trailing edge.

	\begin{figure} [ht]
		\centering
		\includegraphics[width=0.6\columnwidth]{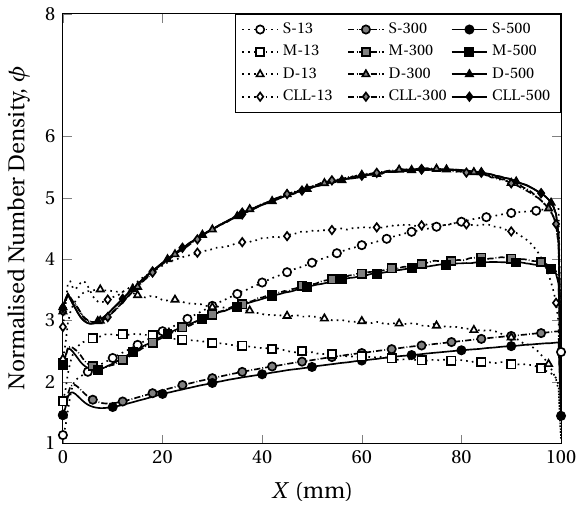}
		\caption{
			\label{fig:numberDensityTemperature}  
			The effect of free-stream temperature and wall model on the normalised number density.   
		}
	\end{figure}
	
	Figure~\ref{fig:numberDensityTemperature} shows the number density along the surface for three free-stream temperatures with four wall models. When the results of the diffuse and CLL wall models are investigated, higher thermal velocities at higher flow temperatures, at 300 $K$ and 500 $K$, increase the frequency of gas-surface interactions, as energetic particles are more likely to hit the surface. This results in a higher number of particles reflecting from the surface, thereby increasing the local number density near the wall. Additionally, for both models, the reflected particles are thermalised near the wall, and thus, the reflected particle distribution becomes nearly identical, leading to similar number densities in both models.
	
	However, the colder gas at 13.32 $K$ interacting with the specular wall has the highest normalised number density in the cold flow case. Since there is not a temperature accommodation or energy exchange with the wall, the velocity magnitude of reflected particles remains unchanged, meaning specular reflections do not randomise and thermalise particle velocities. Thus, cold gas particles with low thermal velocity are slow-moving, which increases the likelihood of them interacting with the surface. The Maxwellian wall model presents similar results to the diffuse wall model, as the Maxwellian model also executes a selection process between the specular and diffuse wall models, and includes thermalisation parameters, which dominates the flow behaviour. 
	
	\subsubsection{Velocity profile}
\begin{figure}[h!]
		\centering
		\begin{subfigure}{0.65\textwidth}
			\includegraphics[width=\textwidth]{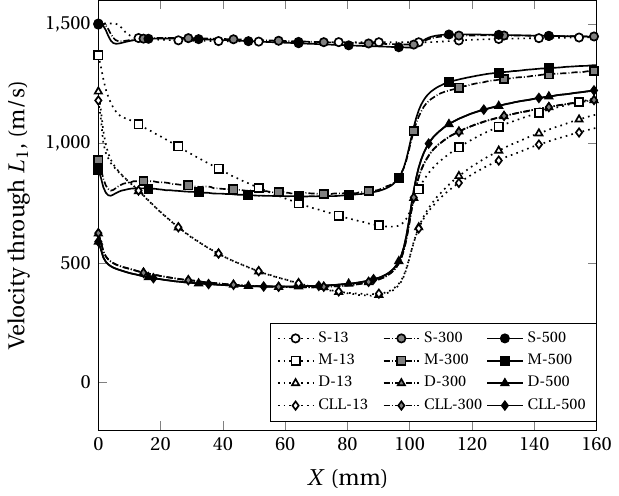}
			\caption{Velocity patterns along the horizontal data collection line.}
			\label{fig:L1VelTemp}
		\end{subfigure}
		\hfill
		\begin{subfigure}{0.65\textwidth}
			\includegraphics[width=\textwidth]{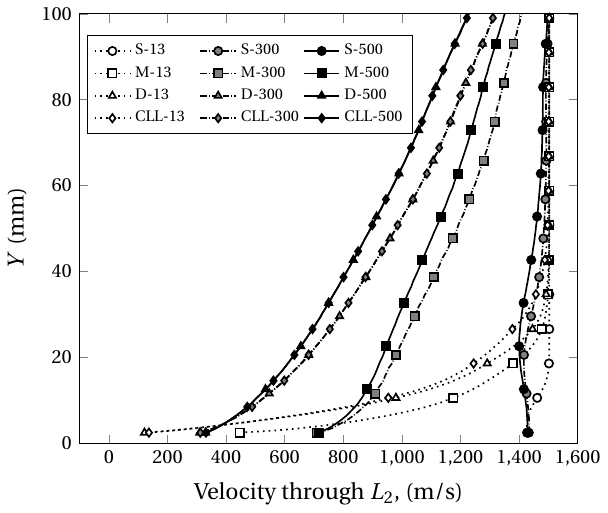}
			\caption{Velocity patterns along the vertical data collection line.}
			\label{fig:L2VelTemp}
		\end{subfigure}
		\caption{Velocity profiles of the specular, Maxwellian, diffuse, and CLL wall models cases
			at the free-stream temperatures of 13.32 $K$, 300 $K$, and 500 $K$ with 15$^{\circ}$ AoA along the $L1$ and $L2$ data collection lines. The wall models are represented by their initials in the legend, along with the corresponding free-stream temperatures.} 
		\label{fig:velocityTemperature}
	\end{figure}

	Figure~\ref{fig:velocityTemperature} presents the velocity profiles for three free-stream temperatures with four wall models. When the results of the specular wall model along the $L1$ and $L2$ data collection lines are analysed, it is seen that an increase in flow temperature has minimal effect on the velocity behaviour near the surface, as the wall model is not capable of thermalising the flow. However, with the rise in temperature, a slight disturbance in the flow can be observed along the vertical axis. 
	
	The effect of increased flow temperature on the flow velocity exhibits different behaviour in fully diffuse and fully accommodated walls. As previously discussed, these active walls slow the flow more than specular walls due to energy exchange mechanisms. In the colder flow case, the flow velocity gradually decreases along the surface, from the leading edge to the trailing edge. This behaviour can be attributed to the low thermal velocity and energy of particles in colder and denser flows, in addition cumulative gas-surface interactions, cause a gradual deceleration. Thus, lower initial thermal velocity in colder flows result in gradual deceleration as energy and momentum exchange occur over a longer distance along the plate with active wall models. Conversely, in less dense and high-energy flows, particles possess higher momentum. This leads to a more constant velocity distribution near the surface. Consequently, high thermal velocity and momentum in warmer flows cause rapid deceleration due to frequent and efficient momentum exchange with the surface. This rapid stabilisation occurs because the particles, due to their high velocity and momentum, quickly adjust to the wall's influence and are less affected over longer distances. 
	
	The Maxwellian wall model with a diffuse fraction of 0.5 demonstrates the effect of partial diffusivity on flow velocity, exhibiting moderate behaviour. In the cold flow case, it mimics the gradual deceleration of the diffuse wall but slows down less because the specular fraction allows some particles to retain their velocities. In warmer flows, the velocity remains nearly constant along the surface, similar to fully diffuse and CLL walls. 
	
\pagebreak 

	\subsection{Free-stream Velocity} \label{sec:velocity} 
	A comparative analysis is conducted using the results of simulation groups 3, 4, and 5 to investigate the effect of changes in the free-stream velocity, $U_{\infty}$. The free-stream temperature, $T_{\infty}$, and AoA are maintained at 500 $K$ and 15$^{\circ}$, respectively, and three free-stream velocities are chosen: 1503 $m/s$, 4500 $m/s$, and 7500 $m/s$. The molecular number density, $n_{\infty}$, is calculated as 9.902 $\times$ $10^{18}$ $1/m^{3}$, corresponding to 1.237 $\times$ $10^{8}$ DSMC simulator particles. For varying flow velocities, the cell resilience time changes; therefore, the time steps for simulation groups 3, 4, and 5 are chosen as 2 $\times$ $10^{-7}$ s, 9 $\times$ $10^{-8}$ s, and 5 $\times$ $10^{-8}$ s, respectively, to correctly capture the intermolecular collisions. Since the pressure and temperature of the flow remain constant in these three cases, the rarefaction level is also maintained with a Kn of 1.5, which places the flow in the transition regime. 
	
	\subsubsection{Scattering} 
\begin{figure}[h!]
		\centering
		\includegraphics[width=0.6\columnwidth]{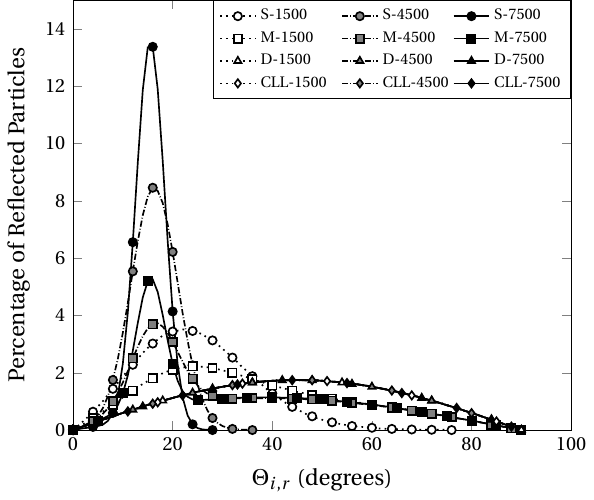}
		\caption{Interaction angles vs percentage of reflected particles from overall surface of the wall for three different free-stream velocities and four various wall models. S, M, D, and CLL represent specular, Maxwellian, diffuse, and CLL wall models. 1500, 4500, and 7500 represent the free-stream velocities.} 
		\label{fig:scatteringVelocity}
	\end{figure}

	Figure~\ref{fig:scatteringVelocity} presents the reflection angle vs percentage of the particles at three different free-stream velocities. 
	
	\begin{figure}[h!]
		\centering
		\begin{subfigure}{0.5\textwidth}
			\includegraphics[width=\textwidth]{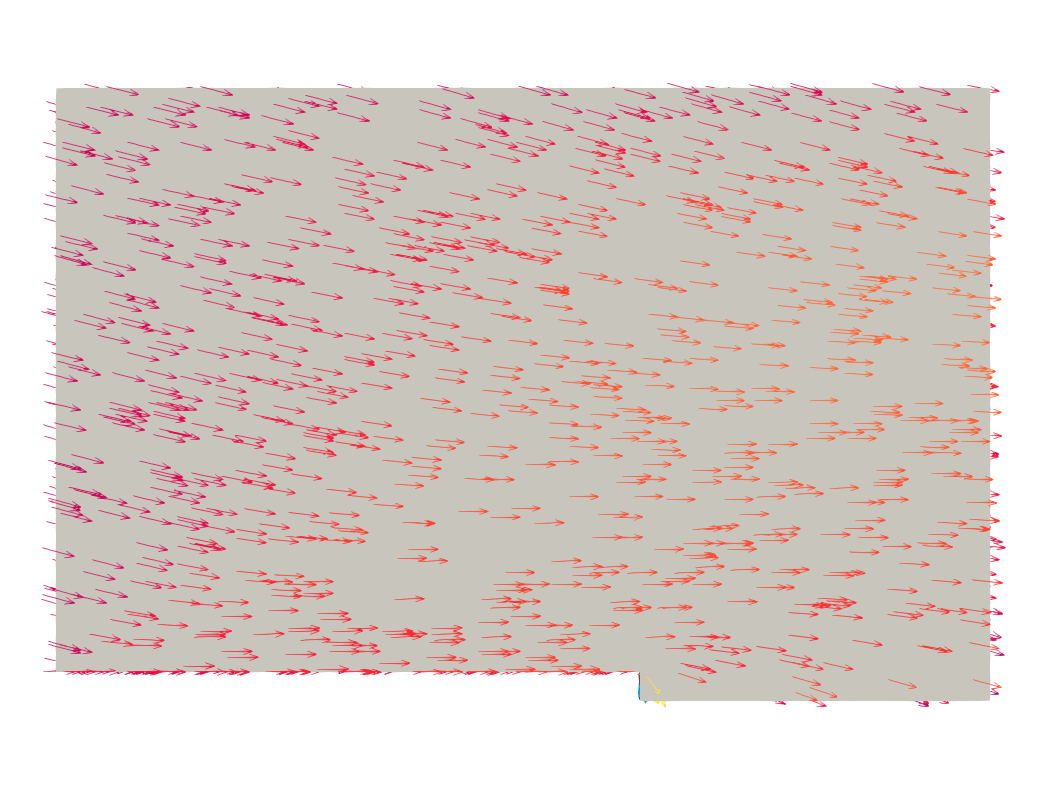}
			\caption{$U_{\infty}$ = 1503 $m/s$, Specular.}
			\label{fig:gylphVel-S-0}
		\end{subfigure}
		\hfill
		\begin{subfigure}{0.5\textwidth}
			\includegraphics[width=\textwidth]{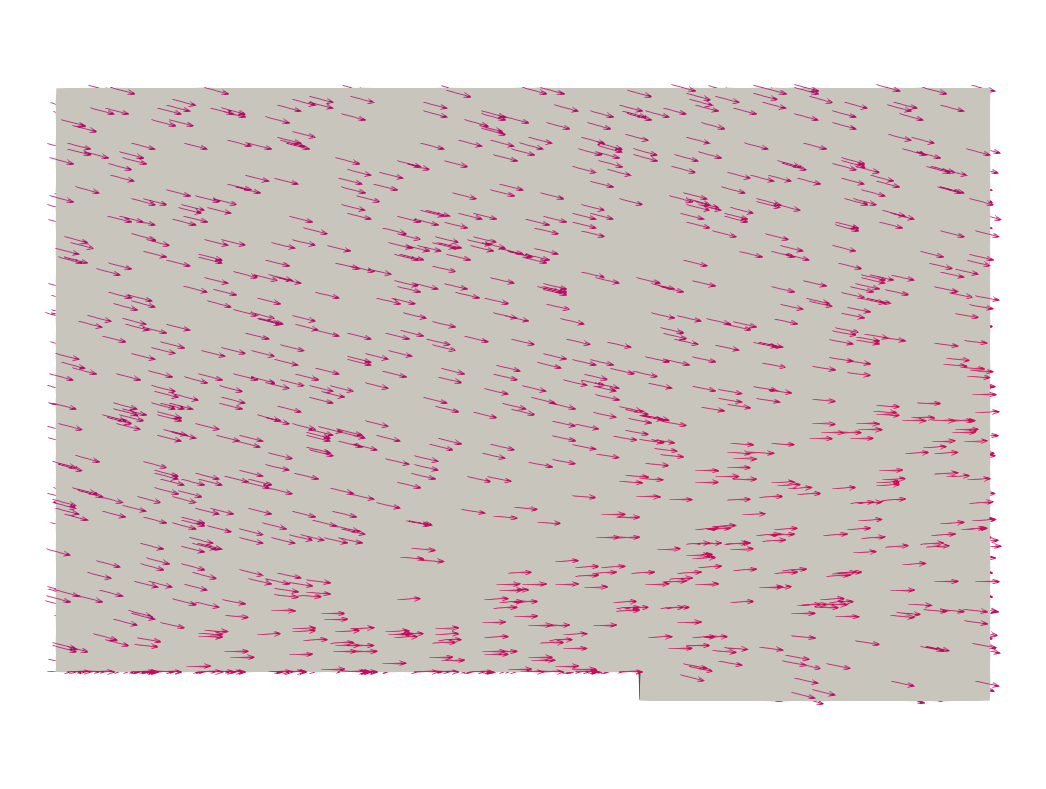}
			\caption{$U_{\infty}$ = 4500 $m/s$, Specular.}
			\label{fig:gylphVel-S-15}
		\end{subfigure}
		\hfill
		\begin{subfigure}{0.5\textwidth}
			\includegraphics[width=\textwidth]{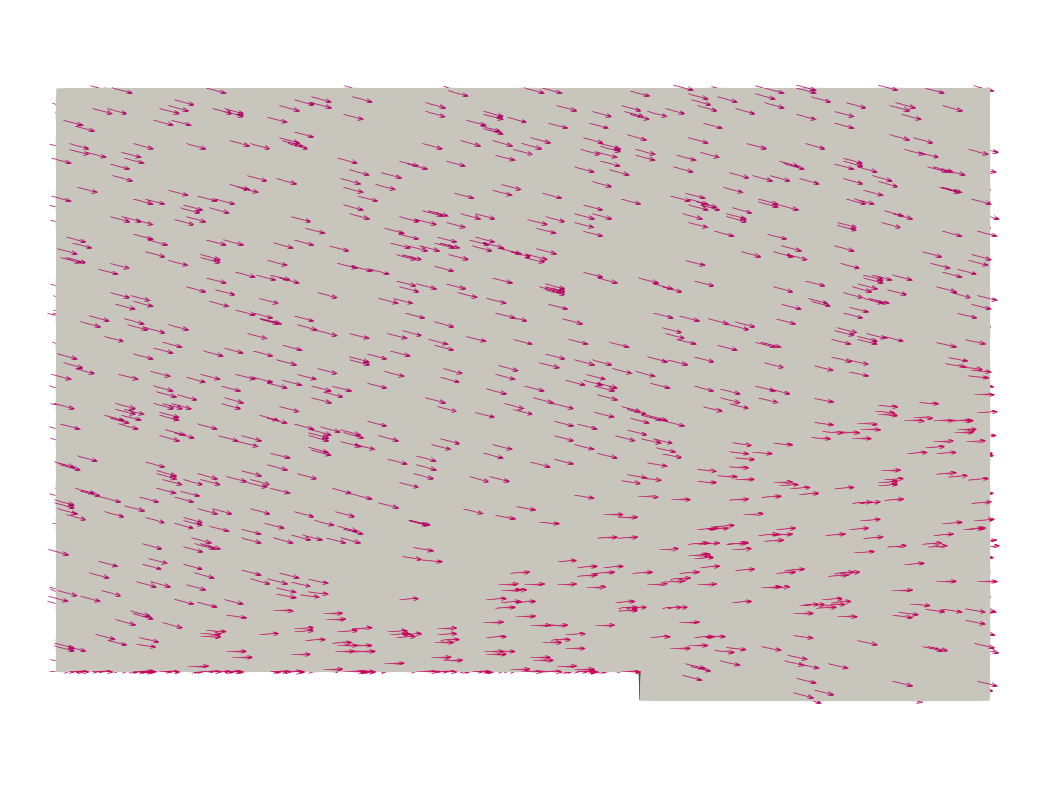}
			\caption{$U_{\infty}$ = 7500 $m/s$, Specular.}
			\label{fig:gylphVel-S-30}
		\end{subfigure}
		\caption{Flow visualisation of the specular wall model cases for all three free-stream velocities.} 
		\label{fig:gylphVelocity-S}
	\end{figure} 
	
	As the reflection behaviours of the particles at 1500 $m/s$ and 500 $K$ are discussed in \S~\ref{sec:temperature}, this relatively lower speed and higher rarefaction level create a reflection trend with 3.51\% of particles reflected 21$^{\circ}$ in the specular wall model. However, the increment in the free-stream velocity dominates the particle interactions, so 8.45\% and 13.38\% of particles at 4500 $m/s$ and 7500 $m/s$, respectively, reflect with 16$^{\circ}$. This trend can be also seen in the Maxwellian wall model, as the flow at 1500 $m/s$ has a more distributed reflection peak with 2.22\% of particles reflecting with between 22$^{\circ}$-26$^{\circ}$. However, 3.73\% and 5.26\% of particles reflect with 17$^{\circ}$ and 16$^{\circ}$, and creates the peaks. The formation of the reflection flow is visualised in Figure~\ref{fig:gylphVelocity-S} for the specular wall model when the velocity of the free-stream flow is increased. In this case too, the change in velocity does not affect the reflection angle in the diffuse and CLL wall models due to the fully randomisation. 
	
	\subsubsection{Surface parameters}
\begin{figure} [ht]
		\centering
		\includegraphics[width=0.6\columnwidth]{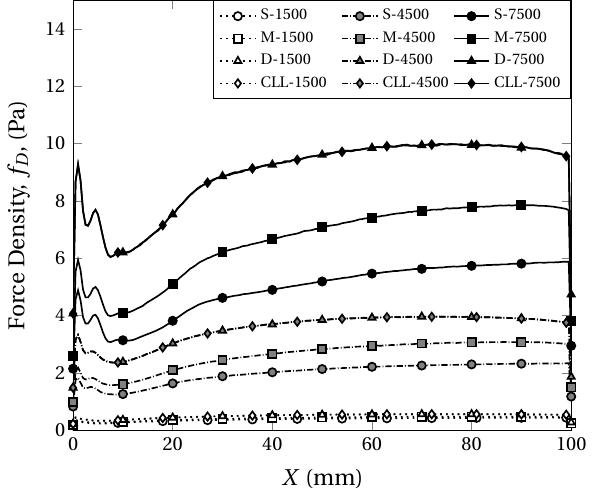}
		\caption{
			\label{fig:fdVelocity}  
			The effect of free-stream velocity and wall model on the force density.    
		}
	\end{figure}

	Figure~\ref{fig:fdVelocity} shows the effect of free-stream velocity and wall model on force density. When the velocity groups are analysed individually, it is observed that higher surface forces are present in the active walls. Additionally, as the flow accelerates, the surface forces increase for all wall models. 
	
	When the free-stream velocity increases, the momentum flux to the surface increases by the square of the velocity. This greater momentum leads to stronger particle-wall interactions, therefore, higher surface forces occur. Additionally, in cases where the flow strikes through the surface, such as here with a 15$^{\circ}$ AoA, high velocity particles approach the surface stronger. This increases the rate of momentum transfer to the surface. Near the trailing edge, the cumulative interaction grows due to the continued interaction of particles along the surface. Consequently, higher velocity creates a greater impact on the surface in terms of forces. Considering the wall models and energy exchange previously discussed, it is observed that active walls predict higher surface forces, as expected. 
	
\begin{figure} [ht]
		\centering
		\includegraphics[width=0.6\columnwidth]{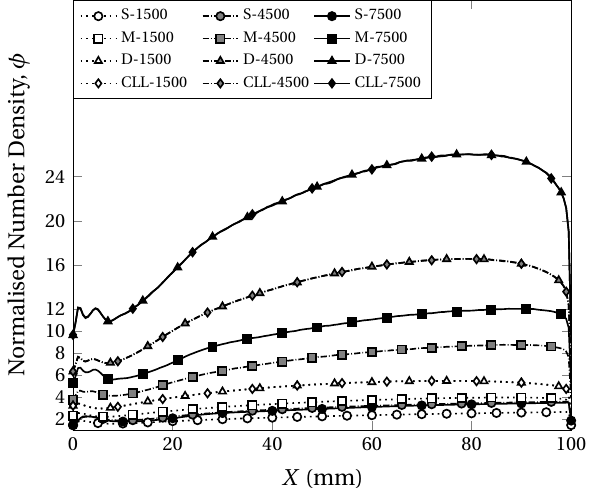}
		\caption{
			\label{fig:numberDensityVelocity}  
			The effect of free-stream velocity and wall model on the normalised number density.  
		}
	\end{figure}

	Figure~\ref{fig:numberDensityVelocity} shows the ratio of the number density on the surface to the free-stream number density. At higher velocities, particles hit the surface more frequently due to their higher normal velocity component, which increases the interaction with the surface. The tangential component causes particles to `sweep' along the surface, increasing the particle density near the trailing edge. This creates a localised buildup of reflected particles, leading to an increase in surface number density. While higher flow velocities increase the number density on the surface, active walls also contribute to the surface density. 
	
	\subsubsection{Velocity profile} 
\begin{figure}[h!]
		\centering
		\begin{subfigure}{0.6\textwidth}
			\includegraphics[width=\textwidth]{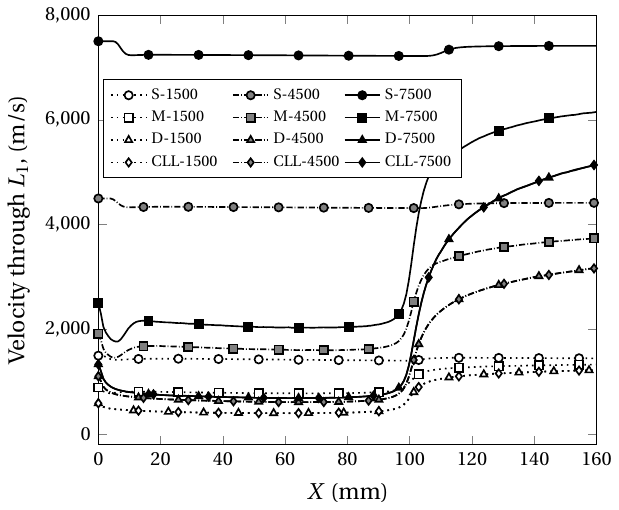}
			\caption{Velocity patterns along the horizontal data collection line.}
			\label{fig:L1VelVel}
		\end{subfigure}
		\hfill
		\begin{subfigure}{0.6\textwidth}
			\includegraphics[width=\textwidth]{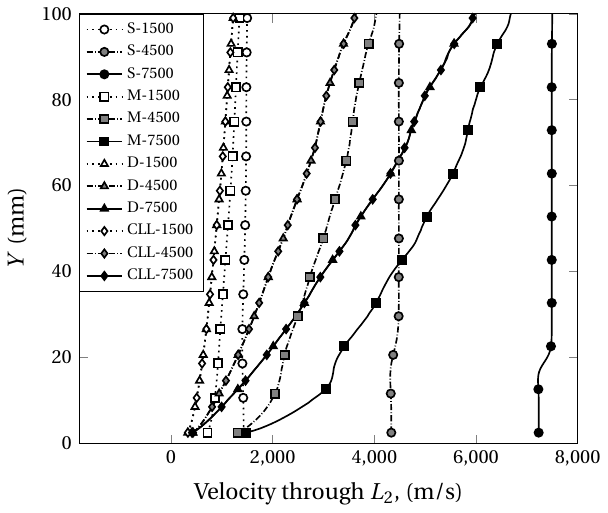}
			\caption{Velocity patterns along the vertical data collection line.}
			\label{fig:L2VelVel}
		\end{subfigure}
		\caption{Velocity profiles of the specular, Maxwellian, diffuse, and CLL wall models cases
			at the free-stream velocities of 1503 $m/s$, 4500 $m/s$, and 7500 $m/s$ with 15$^{\circ}$ AoA along the $L1$ and $L2$ data collection lines. The wall models are represented by their initials in the legend, along with the corresponding free-stream velocities.} 
		\label{fig:velocityVelocity}
	\end{figure}

	Figure~\ref{fig:velocityVelocity} shows the velocity profile along the data collection lines $L1$ and $L2$. Horizontally, higher flow disturbances are observed for the fully active walls compared to the specular and Maxwellian walls. Furthermore, it is measured that the flow just above the surface reaches its minimum in all three cases with different free-stream velocities when active wall models are employed. However, when the flow leaves the surface -after the trailing edge, the velocity gradually returns to almost its free-stream value before exiting the control volume. When the vertical velocity profile is investigated, it is observed that flow disturbance increases with higher free-stream velocities. Additionally, the effect of the diffuse/accommodation properties of the wall on the velocity profile can be deduced from the presented data on $L2$. 
	
        \pagebreak 
	\subsection{Species}
	All previous simulations were conducted with pure N$_{2}$ flow. A comparison was performed to examine the differences between simulations using pure N$_{2}$ and a mixture of N$_{2}$ and O$_{2}$ as the working gas. Additionally, to approximate the conditions of VLEO applications while maintaining the structure of the parametric study, the free-stream temperature and velocity were set to 700 $K$ and 7500 $m/s$, respectively. The AoA is set to 15$^{\circ}$, as in previous simulation groups. To connect the findings of previous studies with this comparison, simulation group 5 is used as a look-up reference. Therefore, in this section, simulation groups 5, 6, and 7 are analysed to investigate changes in flow and surface parameters based on species composition. The initial conditions for simulation group 5 are provided in \S~\ref{sec:velocity}. The molecular number density, $n_{\infty}$, for simulation groups 6 and 7 is the same, calculated as 7.07 $\times$ $10^{18}$ $1/m^{3}$, corresponding to 8.84 $\times$ $10^{7}$ DSMC simulators. While group 6 involves pure N$_{2}$ flow, the gas flow in group 7 consists of a mixture of 79\% N$_{2}$ and 21\% O$_{2}$. It should be emphasised that this ratio was not selected to represent a specific altitude but rather to illustrate the effects of different species. 
	
	\subsubsection{Scattering} 
	Figure~\ref{fig:scatteringSpecies} shows the interaction angles of particles with the surface at increasing free-stream temperatures for pure N$_{2}$ and N$_{2}$-O$_{2}$ mixture flows. 
	
	\begin{figure}[h!]
		\centering
		\includegraphics[width=0.6\columnwidth]{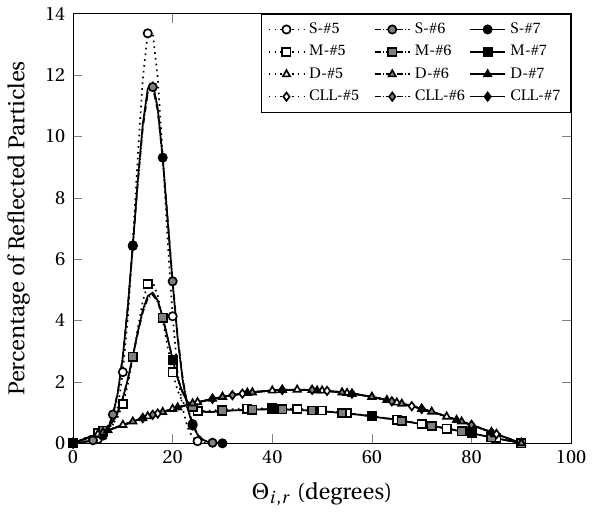}
		\label{fig:scatteringSpecies}
		\caption{Interaction angles vs percentage of reflected particles from overall surface of the wall for pure nitrogen and mixture gas flows and four various wall models. S, M, D, and CLL represent specular, Maxwellian, diffuse, and CLL wall models. 5, 6, and 7 represent the simulation groups.} 
		\label{fig:scatteringSpecies}
	\end{figure}
		
	For the specular wall model, the free-stream temperature affects the scattering profile, as discussed in \S~\ref{sec:temperature}. However, pure and gas mixture flows show the same scattering trend at the same free-stream temperature in both specular and active wall models. This is because only the pre-collision velocity vectors are significant in the specular model, while the scattering process in active walls is conducted by randomisation using the wall temperature, which is maintained in all cases. 
	
	\subsubsection{Surface parameters} As the effect of the change in the free-stream temperature is previously discussed in \S~\ref{sec:temperature}, the pure and mixture flows are compared in this section. The simulation groups of 6 and 7 represent the pure and gas mixture at 700 $K$, respectively. 
	
	\begin{figure} [ht]
		\centering
		\includegraphics[width=0.6\columnwidth]{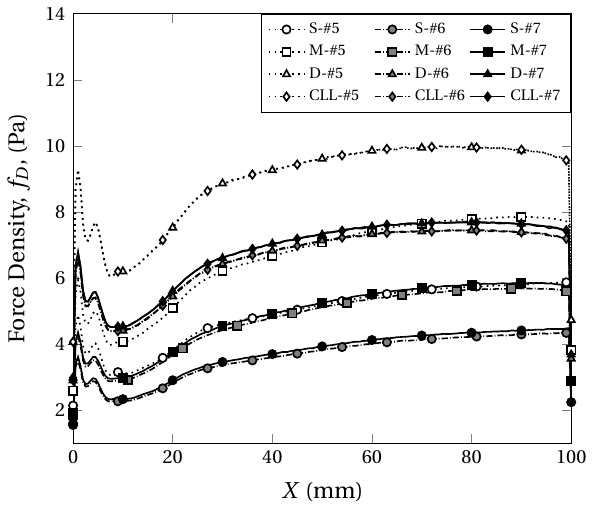}
		\caption{
			\label{fig:fdSpecies}  
			The effect of species variations in the flow, free-stream temperature, and wall model on force density. 
		}
	\end{figure} 
	
	Figure~\ref{fig:fdSpecies} shows the variations in the surface forces. It is observed that the gas mixture generates slightly higher surface forces compared to pure N$_{2}$ flow, e.g., 2.75\% more force at $X$ = 20 $mm$ in the CLL wall model. In this case, the flow properties are quite similar for pure and mixture flows, and the wall temperature is maintained. However, an O$_{2}$ molecule in the VHS particle model is 6.6 $\times$ 10$^{-27}$ $kg$ heavier than a N$_{2}$ molecule, where 21\% of the mixture is O$_{2}$~\cite{bird}. 
	
\begin{figure} [ht]
		\centering
		\includegraphics[width=0.6\columnwidth]{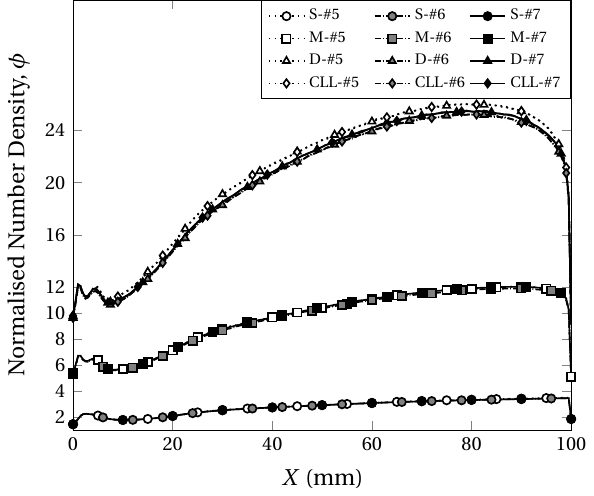}
		\caption{
			\label{fig:numberDensitySpecies}  
			The effect of species variations in the flow, free-stream temperature, and wall model on normalised number density. 
		}
	\end{figure}

	This increase in molecular mass also affects the number density at the surface. Since the heavier molecules move slower on average at the same temperature, i.e. $v_{rms} = \sqrt{\sfrac{3kT}{m}}$, where $v_{rms}$ is the root-mean-square velocity and $m$ is the molecular mass. An additional contribution from the wall model is the diffuse fraction/accommodation coefficient, as O$_{2}$ and N$_{2}$ molecules may exhibit different scattering behaviours. O$_{2}$’s greater mass might lead to more energy exchange compared to N$_{2}$, further contributing to localised number density variations near the wall. Figure~\ref{fig:numberDensitySpecies} shows the distribution of number density along the surface. 
	
	\subsubsection{Velocity profile} 
\begin{figure}[h!]
		\centering
		\begin{subfigure}{0.6\textwidth}
			\includegraphics[width=\textwidth]{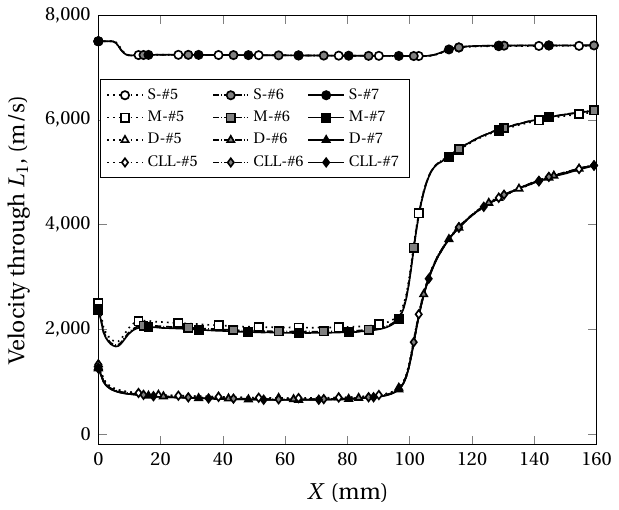}
			\caption{Velocity patterns along the horizontal data collection line.}
			\label{fig:L1VelSpecies}
		\end{subfigure}
		\hfill
		\begin{subfigure}{0.6\textwidth}
			\includegraphics[width=\textwidth]{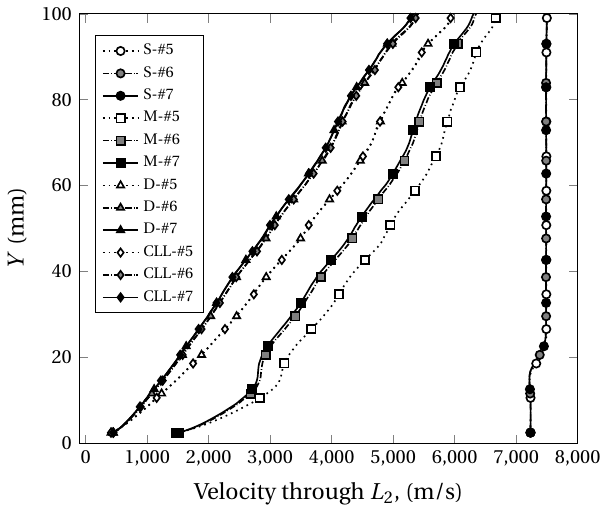}
			\caption{Velocity patterns along the vertical data collection line.}
			\label{fig:L2VelSpecies}
		\end{subfigure}
		\caption{Velocity profiles of the specular, Maxwellian, diffuse, and CLL wall models cases with pure and mixture flows at 15$^{\circ}$ AoA along the $L1$ and $L2$ data collection lines. The wall models are represented by their initials in the legend, along with the corresponding the simulation groups.} 
		\label{fig:velocitySpecies}
	\end{figure}

Figure~\ref{fig:velocitySpecies} shows the change in velocity with respect to the increase in free-stream temperature and change in species of the flow. On the horizontal measurement line, the velocities in all three cases are almost identical when the specular wall model is chosen. However, the gas mixture is slightly slower, e.g., ~1.5\% at the $X$ = 20 $mm$ location in the CLL wall model case, compared to the pure flow. This can also be observed through the vertical data collection line, where the gas mixture shows a slightly slower trend compared to the pure gas flow simulations. 
	
\pagebreak 
    
	\section{Discussion}
	This study presents parametric research to determine how changes in flow parameters and surface models affect the modeling of gas-surface interactions and the prediction of flow near the wall and in the control volume. Furthermore, the capabilities of the default boundary conditions of dsmcFoam were also investigated. 
    
    Understanding these interactions is crucial for ABEP applications, particularly in terms of surface force formation, as well as propellant collection and compression processes. Therefore, the numerical gas-surface interaction model is significant in predicting the flow and interaction parameters. Consequently, this research aims to deepen our understanding of gas-surface interactions under varying conditions with a comparison of boundary conditions in dsmcFoam to inform the development of a new boundary condition that better represents realistic gas-surface interaction phenomena.
	
	In this study, the effects of AoA, free-stream temperature and velocity, and pure versus gas mixture simulations were analysed using the four default boundary conditions in dsmcFoam, which are specular, Maxwellian, diffuse, and CLL boundary conditions.
	
	\subsection{Flow Parameters} 
	For scattering behaviours, while the AoA, free-stream temperature, and velocity significantly impact reflections by the specular wall model and have a slight inherent effect on the Maxwellian wall. These parameters do not influence the scattering behaviour of the fully active walls as they randomise scattering. However, due to variations in post-collision calculations among wall models, the formation of the reflected flow in the control volume changes. Flow disturbance increases with higher diffuse fraction/accommodation coefficients, with the CLL model creating more disturbed flow around the wall compared to other models. 
	
	Increasing flow temperature results in a more rarefied flow when pressure is maintained across all cases. Surface forces decrease, whereas the ratio of number density to free-stream number density, which is normalised number density, increases compared to colder flows. Conversely, increasing flow velocity raises both surface forces and the normalised number density. The flow velocity near the wall is also influenced by AoA, free-stream temperature, and velocity. Higher AoA and reduced flow temperature or velocity led to a slower region near the wall. While flow parameters significantly impact gas-surface interactions, the contributions of numerical wall models are also notable.
	
	\subsection{Wall Models} 
	Maxwellian and diffuse wall models do not account for pre-collision terms, whereas the CLL wall model creates a reflection kernel that incorporates both pre- and post-collision parameters. The Maxwellian and diffuse wall models employ the same energy update approach—based on the equipartition theorem—since the Maxwellian model is a combination of specular and diffuse models. The equipartition theorem states that, for a system in thermal equilibrium, each degree of freedom (e.g., translational, rotational, vibrational) of a particle contributes an equal amount to the total energy. The CLL wall model requires user-defined normal, tangential, and rotational accommodation coefficients for its velocity and energy updates, where the rotational energy exchange at the surface is performed. The current version of this boundary condition in dsmcFoam does not update the vibrational energy. The specular wall model in dsmcFoam does not update the energy levels during the post-collision procedure; it solely provides a perfectly elastic collision by altering the direction of the normal component of the velocity vector of particles. 
	
	The effects of wall models can be observed in the scattering kernel, flow patterns near the wall, and surface parameters. As mentioned earlier, while scattering angles are not functions of flow parameters for active walls (e.g., diffuse and CLL, as well as partially Maxwellian), they have strong effects on simulations using the specular wall model. Different wall models produce distinct flow physics near the wall, with the fully accommodated CLL model creating more disturbed flow, higher surface forces, and a greater normalised number density compared to other models. Additionally, the diffuse wall model could not capture the same trend as the CLL model for surface forces and the normalised number density in the colder flow cases. 
	
	\section{Conclusion} 
    Gas-surface interactions are complex phenomena that require an accurate numerical tool to simultaneously model multiple interaction physics for diverse particle mixtures with varying species. These interactions strongly influence flow characteristics under varying conditions and wall model preferences, as observed in this study. This is particularly significant for ABEP systems, which operate in diverse atmospheric conditions where surface forces play a critical role in the systems' design and operation. However, the default boundary conditions do not adequately capture the characteristic of reflected flow due to the properties of surface material (surface roughness treatment, thermal properties of the surface, constant accommodation coefficients, etc.), evaluating each collision based on the surface's interaction with individual particles, which include various species such as N$_{2}$, O$_{2}$, O, He, Ar, etc. with various macroscopic properties. 
	
	To address this complex physics, a wall model that incorporates both pre- and post-collision parameters with energy updates, while determining the reflection kernel between specular, diffuse, and isotropic reflections, using extensive material properties for a realistic scattering kernel is necessary. Such a model would improve predictions of gas-surface interactions for multi-species mixtures under varying environmental conditions. Depending on the design and application, an optimal combination of surface coating materials could be applied to different sections of the intake channels and compressor section. 

\bmhead{Acknowledgements}
The authors would like to acknowledge the use of the Computational Shared Facility (CSF) at the University of Manchester.

\section*{Declarations}

\begin{itemize}
\item \textbf{Funding} This research was supported by the UK Space Agency's National Space Innovation Programme (NSIP) under the grant number UKSAG23-0092-008. 
\item \textbf{Conflict of interest/Competing interests} No conflict of interest. 
\item \textbf{Ethics approval and consent to participate} N/A. 
\item \textbf{Consent for publication - Open Access} This article is licensed under a Creative Commons Attribution 4.0 International License, which permits use, sharing, adaptation, distribution and reproduction in any medium or format, as long as you give appropriate credit to the original author(s) and the source, provide a link to the Creative Commons licence, and indicate if changes were made. The images or other third party material in this article are included in the article's Creative Commons licence, unless indicated otherwise in a credit line to the material. If material is not included in the article's Creative Commons licence and your intended use is not permitted by statutory regulation or exceeds the permitted use, you will need to obtain permission directly from the copyright holder. To view a copy of this licence, visit http://creativecommons.org/licenses/by/4.0/. 
\item \textbf{Data availability} The simulation data are available upon request. Authorisation from the funding entity or project partners is required to access this data. 
\item \textbf{Materials availability} N/A. 
\item \textbf{Code availability} The code used to perform simulations is the default dsmcFoam. 
\item \textbf{Author contribution} \textbf{M.B.A.:} Writing – Original draft preparation, Simulation, Data curation and Visualisation. \textbf{N.H.C.:} Supervision, Reviewing and Editing. \textbf{K.L.S.:} Supervision, Reviewing and Editing. \textbf{P.C.E.R.:} Supervision, Reviewing and Editing. \textbf{M.N.:} Reviewing. \textbf{M.G.:} Reviewing. \textbf{S.V.:} Reviewing. 
\end{itemize}

\bibliography{sn-bibliography}


\begin{thebibliography}{23}
\ifx \bisbn   \undefined \def \bisbn  #1{ISBN #1}\fi
\ifx \binits  \undefined \def \binits#1{#1}\fi
\ifx \bauthor  \undefined \def \bauthor#1{#1}\fi
\ifx \batitle  \undefined \def \batitle#1{#1}\fi
\ifx \bjtitle  \undefined \def \bjtitle#1{#1}\fi
\ifx \bvolume  \undefined \def \bvolume#1{\textbf{#1}}\fi
\ifx \byear  \undefined \def \byear#1{#1}\fi
\ifx \bissue  \undefined \def \bissue#1{#1}\fi
\ifx \bfpage  \undefined \def \bfpage#1{#1}\fi
\ifx \blpage  \undefined \def \blpage #1{#1}\fi
\ifx \burl  \undefined \def \burl#1{\textsf{#1}}\fi
\ifx \doiurl  \undefined \def \doiurl#1{\url{https://doi.org/#1}}\fi
\ifx \betal  \undefined \def \betal{\textit{et al.}}\fi
\ifx \binstitute  \undefined \def \binstitute#1{#1}\fi
\ifx \binstitutionaled  \undefined \def \binstitutionaled#1{#1}\fi
\ifx \bctitle  \undefined \def \bctitle#1{#1}\fi
\ifx \beditor  \undefined \def \beditor#1{#1}\fi
\ifx \bpublisher  \undefined \def \bpublisher#1{#1}\fi
\ifx \bbtitle  \undefined \def \bbtitle#1{#1}\fi
\ifx \bedition  \undefined \def \bedition#1{#1}\fi
\ifx \bseriesno  \undefined \def \bseriesno#1{#1}\fi
\ifx \blocation  \undefined \def \blocation#1{#1}\fi
\ifx \bsertitle  \undefined \def \bsertitle#1{#1}\fi
\ifx \bsnm \undefined \def \bsnm#1{#1}\fi
\ifx \bsuffix \undefined \def \bsuffix#1{#1}\fi
\ifx \bparticle \undefined \def \bparticle#1{#1}\fi
\ifx \barticle \undefined \def \barticle#1{#1}\fi
\bibcommenthead
\ifx \bconfdate \undefined \def \bconfdate #1{#1}\fi
\ifx \botherref \undefined \def \botherref #1{#1}\fi
\ifx \url \undefined \def \url#1{\textsf{#1}}\fi
\ifx \bchapter \undefined \def \bchapter#1{#1}\fi
\ifx \bbook \undefined \def \bbook#1{#1}\fi
\ifx \bcomment \undefined \def \bcomment#1{#1}\fi
\ifx \oauthor \undefined \def \oauthor#1{#1}\fi
\ifx \citeauthoryear \undefined \def \citeauthoryear#1{#1}\fi
\ifx \endbibitem  \undefined \def \endbibitem {}\fi
\ifx \bconflocation  \undefined \def \bconflocation#1{#1}\fi
\ifx \arxivurl  \undefined \def \arxivurl#1{\textsf{#1}}\fi
\csname PreBibitemsHook\endcsname

\bibitem[\protect\citeauthoryear{Rapisarda et~al.}{2023}]{rapisarda2023design}
\begin{barticle}
\bauthor{\bsnm{Rapisarda}, \binits{C.}},
\bauthor{\bsnm{Roberts}, \binits{P.C.E.}},
\bauthor{\bsnm{Smith}, \binits{K.L.}}:
\batitle{Design and optimisation of a passive {A}tmosphere-{B}reathing
  {E}lectric {P}ropulsion ({ABEP}) intake}.
\bjtitle{Acta Astronaut.}
\bvolume{202},
\bfpage{77}--\blpage{93}
(\byear{2023})
\doiurl{10.1016/j.actaastro.2022.09.047}
\end{barticle}
\endbibitem

\bibitem[\protect\citeauthoryear{Rapisarda}{2023}]{rapisarda2023modelling}
\begin{barticle}
\bauthor{\bsnm{Rapisarda}, \binits{C.}}:
\batitle{Modelling and simulation of atmosphere-breathing electric propulsion
  intakes via direct simulation {M}onte {C}arlo: {A} study of the air-breathing
  ion engine}.
\bjtitle{CEAS Space J.}
\bvolume{15}(\bissue{2}),
\bfpage{357}--\blpage{370}
(\byear{2023})
\end{barticle}
\endbibitem

\bibitem[\protect\citeauthoryear{Andrews et~al.}{2024}]{andrews2024cathode}
\begin{barticle}
\bauthor{\bsnm{Andrews}, \binits{S.}},
\bauthor{\bsnm{Andriulli}, \binits{R.}},
\bauthor{\bsnm{Souhair}, \binits{N.}},
\bauthor{\bsnm{Magarotto}, \binits{M.}},
\bauthor{\bsnm{Ponti}, \binits{F.}}:
\batitle{Cathode-less {RF} plasma thruster design and optimisation for an
  atmosphere-breathing electric propulsion ({ABEP}) system}.
\bjtitle{Acta Astronaut.}
\bvolume{225},
\bfpage{833}--\blpage{844}
(\byear{2024})
\doiurl{10.1016/j.actaastro.2024.09.041}
\end{barticle}
\endbibitem

\bibitem[\protect\citeauthoryear{Romano et~al.}{2021}]{romano2021intake}
\begin{barticle}
\bauthor{\bsnm{Romano}, \binits{F.}},
\bauthor{\bsnm{Espinosa-Orozco}, \binits{J.}},
\bauthor{\bsnm{Pfeiffer}, \binits{M.}},
\bauthor{\bsnm{Herdrich}, \binits{G.}},
\bauthor{\bsnm{Crisp}, \binits{N.H.}},
\bauthor{\bsnm{Roberts}, \binits{P.}},
\bauthor{\bsnm{Holmes}, \binits{B.}},
\bauthor{\bsnm{Edmondson}, \binits{S.}},
\bauthor{\bsnm{Haigh}, \binits{S.}},
\bauthor{\bsnm{Livadiotti}, \binits{S.}},
\bauthor{\bsnm{Macario-Rojas}, \binits{A.}},
\bauthor{\bsnm{Oiko}, \binits{V.T.A.}},
\bauthor{\bsnm{Sinpetru}, \binits{L.A.}},
\bauthor{\bsnm{Smith}, \binits{K.}},
\bauthor{\bsnm{Becedas}, \binits{J.}},
\bauthor{\bsnm{Sulliotti-Linner}, \binits{V.}},
\bauthor{\bsnm{Bisgaard}, \binits{M.}},
\bauthor{\bsnm{Christensen}, \binits{S.}},
\bauthor{\bsnm{Hanessian}, \binits{V.}},
\bauthor{\bsnm{Kauffman~Jensen}, \binits{T.}},
\bauthor{\bsnm{Nielsen}, \binits{J.}},
\bauthor{\bsnm{Chan}, \binits{Y.-A.}},
\bauthor{\bsnm{Fasoulas}, \binits{S.}},
\bauthor{\bsnm{Traub}, \binits{C.}},
\bauthor{\bsnm{Garc\'{i}a-Almi\~{n}ana}, \binits{D.}},
\bauthor{\bsnm{Rodr\'{i}guez-Donaire}, \binits{S.}},
\bauthor{\bsnm{Sureda}, \binits{M.}},
\bauthor{\bsnm{Kataria}, \binits{D.}},
\bauthor{\bsnm{Belkouchi}, \binits{B.}},
\bauthor{\bsnm{Conte}, \binits{A.}},
\bauthor{\bsnm{Seminari}, \binits{S.}},
\bauthor{\bsnm{Villain}, \binits{R.}}:
\batitle{Intake design for an atmosphere-breathing electric propulsion system
  ({ABEP})}.
\bjtitle{Acta Astronaut.}
\bvolume{187},
\bfpage{225}--\blpage{235}
(\byear{2021})
\doiurl{10.1016/j.actaastro.2021.06.033}
\end{barticle}
\endbibitem

\bibitem[\protect\citeauthoryear{Romano et~al.}{2015}]{romano2015air}
\begin{botherref}
\oauthor{\bsnm{Romano}, \binits{F.}},
\oauthor{\bsnm{Binder}, \binits{T.}},
\oauthor{\bsnm{Herdrich}, \binits{G.}},
\oauthor{\bsnm{Fasoulas}, \binits{S.}},
\oauthor{\bsnm{Sch{\"o}nherr}, \binits{T.}}:
Air-{I}ntake {D}esign {I}nvestigation for an {A}ir-{B}reathing {E}lectric
  {P}ropulsion {S}ystem
(2015).
Paper presented at the Joint Conference of 30th International Symposium on
  Space Technology and Science, and 34th International Electric Propulsion
  Conference and 6th Nano-satellite Symposium Hyogo-Kobe, Japan, 4–10 July
  2015.
\end{botherref}
\endbibitem

\bibitem[\protect\citeauthoryear{Romano et~al.}{2016}]{romano2016intake}
\begin{botherref}
\oauthor{\bsnm{Romano}, \binits{F.}},
\oauthor{\bsnm{Binder}, \binits{T.}},
\oauthor{\bsnm{Herdrich}, \binits{G.}},
\oauthor{\bsnm{Fasoulas}, \binits{S.}},
\oauthor{\bsnm{Sch\"{o}nherr}, \binits{T.}}:
{INTAKE DESIGN FOR AN ATMOSPHERE-BREATHING ELECTRIC PROPULSION SYSTEM}.
Space Propulsion
\textbf{2016}
(2016)
\end{botherref}
\endbibitem

\bibitem[\protect\citeauthoryear{Jackson and
  Marshall}{2018}]{jackson2018conceptual}
\begin{barticle}
\bauthor{\bsnm{Jackson}, \binits{S.W.}},
\bauthor{\bsnm{Marshall}, \binits{R.}}:
\batitle{Conceptual {D}esign of an {A}ir-{B}reathing {E}lectric {T}hruster for
  {C}ube{S}at {A}pplications}.
\bjtitle{J. Spacecr. Rockets}
\bvolume{55}(\bissue{3}),
\bfpage{632}--\blpage{639}
(\byear{2018})
\doiurl{10.2514/1.A33993}
\end{barticle}
\endbibitem

\bibitem[\protect\citeauthoryear{Struchtrup}{2013}]{struchtrup2013maxwell}
\begin{botherref}
\oauthor{\bsnm{Struchtrup}, \binits{H.}}:
Maxwell boundary condition and velocity dependent accommodation coefficient.
Phys. Fluids
\textbf{25}(11)
(2013)
\doiurl{10.1063/1.4829907}
\end{botherref}
\endbibitem

\bibitem[\protect\citeauthoryear{White}{2013}]{white_thesis}
\begin{botherref}
\oauthor{\bsnm{White}, \binits{C.}}:
Benchmarking, development and applications of an open source {DSMC} solver.
PhD thesis
(2013).
Provided by the University of Strathclyde Data System.
\url{https://stax.strath.ac.uk/concern/theses/7m01bk83m?locale=pt-BR}
\end{botherref}
\endbibitem

\bibitem[\protect\citeauthoryear{Bird}{1994}]{bird}
\begin{bbook}
\bauthor{\bsnm{Bird}, \binits{G.A.}}:
\bbtitle{Molecular {G}as {D}ynamics and the {D}irect {S}imulation {M}onte
  {C}arlo of {G}as {F}lows}.
\bpublisher{Clarendon},
\blocation{Oxford}
(\byear{1994})
\end{bbook}
\endbibitem

\bibitem[\protect\citeauthoryear{Gad-el Hak}{1999}]{Gad-el-Hak}
\begin{barticle}
\bauthor{\bsnm{Gad-el-Hak}, \binits{M.}}:
\batitle{The {F}luid {M}echanics of {M}icrodevices—the {F}reeman {S}cholar
  {L}ecture}.
\bjtitle{J. Fluids Eng.}
\bvolume{121}(\bissue{1}),
\bfpage{5}--\blpage{33}
(\byear{1999})
\doiurl{10.1115/1.2822013}
\end{barticle}
\endbibitem

\bibitem[\protect\citeauthoryear{Garcia}{2011}]{garcia}
\begin{botherref}
\oauthor{\bsnm{Garcia}, \binits{A.L.}}:
Direct {S}imulation {M}onte {C}arlo: {T}heory, {M}ethods, and {O}pen
  {C}hallenges,
(2011).
Provided by the NATO Research and Technology Organisation with RTO-EN-AVT-194.
\url{https://apps.dtic.mil/sti/pdfs/ADA582772.pdf}
\end{botherref}
\endbibitem

\bibitem[\protect\citeauthoryear{Bird}{1978}]{bird_annu.rev.}
\begin{barticle}
\bauthor{\bsnm{Bird}, \binits{G.A.}}:
\batitle{{MONTE CARLO SIMULATION OF GAS FLOWS}}.
\bjtitle{Ann. Rev. Fluid Mech.}
\bvolume{10}(\bissue{1}),
\bfpage{11}--\blpage{31}
(\byear{1978})
\doiurl{10.1146/annurev.fl.10.010178.000303}
\end{barticle}
\endbibitem

\bibitem[\protect\citeauthoryear{Oran et~al.}{1998}]{oran}
\begin{barticle}
\bauthor{\bsnm{Oran}, \binits{E.S.}},
\bauthor{\bsnm{Oh}, \binits{C.K.}},
\bauthor{\bsnm{Cybyk}, \binits{B.Z.}}:
\batitle{{DIRECT SIMULATION MONTE CARLO}: {R}ecent {A}dvances and
  {A}pplications}.
\bjtitle{Ann. Rev. Fluid Mech.}
\bvolume{30}(\bissue{1}),
\bfpage{403}--\blpage{441}
(\byear{1998})
\doiurl{10.1146/annurev.fluid.30.1.403}
\end{barticle}
\endbibitem

\bibitem[\protect\citeauthoryear{Scanlon et~al.}{2010}]{scanlon_dsmcFOAM}
\begin{barticle}
\bauthor{\bsnm{Scanlon}, \binits{T.J.}},
\bauthor{\bsnm{Roohi}, \binits{E.}},
\bauthor{\bsnm{White}, \binits{C.}},
\bauthor{\bsnm{Darbandi}, \binits{M.}},
\bauthor{\bsnm{Reese}, \binits{J.M.}}:
\batitle{An open source, parallel {DSMC} code for rarefied gas flows in
  arbitrary geometries}.
\bjtitle{Comput. Fluids}
\bvolume{39}(\bissue{10}),
\bfpage{2078}--\blpage{2089}
(\byear{2010})
\doiurl{10.1016/j.compfluid.2010.07.014}
\end{barticle}
\endbibitem

\bibitem[\protect\citeauthoryear{White et~al.}{2018}]{white}
\begin{barticle}
\bauthor{\bsnm{White}, \binits{C.}},
\bauthor{\bsnm{Borg}, \binits{M.K.}},
\bauthor{\bsnm{Scanlon}, \binits{T.J.}},
\bauthor{\bsnm{Longshaw}, \binits{S.M.}},
\bauthor{\bsnm{John}, \binits{B.}},
\bauthor{\bsnm{Emerson}, \binits{D.R.}},
\bauthor{\bsnm{Reese}, \binits{J.M.}}:
\batitle{dsmc{F}oam+: An {O}pen{FOAM} based direct simulation {M}onte {C}arlo
  solver}.
\bjtitle{Comput. Phys. Commun.}
\bvolume{224},
\bfpage{22}--\blpage{43}
(\byear{2018})
\doiurl{10.1016/j.cpc.2017.09.030}
\end{barticle}
\endbibitem

\bibitem[\protect\citeauthoryear{Maxwell}{1879}]{maxwell1879vii}
\begin{botherref}
\oauthor{\bsnm{Maxwell}, \binits{J.C.}}:
{VII}. {O}n stresses in rarified gases arising from inequalities of
  temperature.
Philos. Trans. R. Soc.
(170),
231--256
(1879)
\doiurl{10.1098/rstl.1879.0067}
\end{botherref}
\endbibitem

\bibitem[\protect\citeauthoryear{Cercignani and
  Lampis}{1971}]{cercignani1971kinetic}
\begin{barticle}
\bauthor{\bsnm{Cercignani}, \binits{C.}},
\bauthor{\bsnm{Lampis}, \binits{M.}}:
\batitle{Kinetic models for gas-surface interactions}.
\bjtitle{Transport {T}heor. {S}tat.}
\bvolume{1}(\bissue{2}),
\bfpage{101}--\blpage{114}
(\byear{1971})
\doiurl{10.1080/00411457108231440}
\end{barticle}
\endbibitem

\bibitem[\protect\citeauthoryear{Lord}{}]{lord1989application}
\begin{botherref}
\oauthor{\bsnm{Lord}, \binits{R.}}:
Application of the {C}ercignani-{L}ampis {S}cattering {K}ernel to {D}irect
  {S}imulation {M}onte {C}arlo {C}alculations.
Paper presented at the 17th International Symposium on Rarefied Gas Dynamics,
  Aachen, Germany, 8-14 July 1990
\end{botherref}
\endbibitem

\bibitem[\protect\citeauthoryear{Lord}{1991}]{lord1991some}
\begin{barticle}
\bauthor{\bsnm{Lord}, \binits{R.G.}}:
\batitle{Some extensions to the {C}ercignani-{L}ampis gas-surface scattering
  kernel}.
\bjtitle{Phys. Fluids A}
\bvolume{3}(\bissue{4}),
\bfpage{706}--\blpage{710}
(\byear{1991})
\doiurl{10.1063/1.858076}
\end{barticle}
\endbibitem

\bibitem[\protect\citeauthoryear{Padilla}{2010}]{padilla2010comparison}
\begin{botherref}
\oauthor{\bsnm{Padilla}, \binits{J.F.}}:
Comparison of {DAC} and {MONACO} {DSMC} {C}odes with {F}lat {P}late
  {S}imulation,
(2010).
Provided by the NTRS - NASA Technical Reports Server with NASA/TM-2010-216835.
\url{https://ntrs.nasa.gov/api/citations/20100029686/downloads/20100029686.pdf}
\end{botherref}
\endbibitem

\bibitem[\protect\citeauthoryear{J.~All\`{e}gre and
  Gottesdiener}{1992}]{allegre}
\begin{barticle}
\bauthor{\bsnm{J.~All\`{e}gre}, \binits{A.C.} \bsuffix{M.~Raffin}},
\bauthor{\bsnm{Gottesdiener}, \binits{L.}}:
\batitle{Rarefied {H}ypersonic {F}low over a {F}lat {P}late with {T}uncated
  {L}eading {E}dge}.
\bjtitle{Prog. Astronaut. Aeronaut.}
\bvolume{160},
\bfpage{285}--\blpage{295}
(\byear{1992})
\end{barticle}
\endbibitem

\bibitem[\protect\citeauthoryear{Agir et~al.}{2022}]{agir2022effect}
\begin{barticle}
\bauthor{\bsnm{Agir}, \binits{M.B.}},
\bauthor{\bsnm{White}, \binits{C.}},
\bauthor{\bsnm{Kontis}, \binits{K.}}:
\batitle{The effect of increasing rarefaction on the formation of {E}dney shock
  interaction patterns: {T}ype-{I} to {T}ype-{VI}}.
\bjtitle{Shock Waves}
\bvolume{32}(\bissue{8}),
\bfpage{733}--\blpage{751}
(\byear{2022})
\end{barticle}
\endbibitem

\end{thebibliography}

\end{document}